\documentclass[a4paper,12pt]{article}
\usepackage{amsmath,amsfonts,epsfig}
\numberwithin{equation}{section}
\usepackage{color}

\def\th{\theta^{\kappa}_{ab}}
\def\ph{\phi^{\kappa}_{ca}}
\def\ps{\psi^{\kappa}_{bc}}

\def\bra{\langle}
\def\ket{\rangle}
\def\d{\mathrm{d}}

\def\tr{\mathrm{tr}}
\def\ap{\alpha^{\prime}}

\def\beq{\begin{equation}}
\def\eeq{\end{equation}}

\newcommand{\ka}{\kappa}

\def\Wh{\hat{W}}

\def\ov{\overline}
\newcommand{\tab}[2]{\theta\left[\begin{array}{c}#1 \\ #2 \end{array}\right]}

\author{K.~Benakli\footnote{kbenakli@lpthe.jussieu.fr} and M.~D.~Goodsell\footnote{goodsell@lpthe.jussieu.fr}}

\date{}
\title{Two-Point Functions of Chiral Fields at One Loop in Type II}

\begin{document}

\maketitle
\vspace{-1cm}
\begin{center}
\emph{Laboratoire de Physique Th\'eorique et Hautes Energies\footnote{Unit\'e Mixte de Recherche du CNRS UMR 7589}, Tour 24-25, 
5eme \'etage, Boite 126,
4 place Jussieu, F-75252 Paris Cedex 05 France}
\end{center}
\abstract{We compute the two-point functions for chiral matter states  in toroidal intersecting D6-brane models. In particular, we provide the techniques to calculate M\"obius strip diagrams including the worldsheet instanton contribution.}

\section{Introduction}

Today's   high energy physics faces   open questions such as the construction of a quantum theory of gravity, explaining baryogenesis, the rotation curves of galaxies, the expansion of the universe, or understanding the gauge  hierarchy problem... These motivate investigations of many extensions of the Standard Model of particle physics, including, for instance, supersymmetric  and higher dimensional ones. The corresponding effective field theories find in string theory  a unique framework for  an  ultraviolet completion.  There, the quantum field theory peturbative expansion is replaced by a sum over  world-sheet surfaces, each order being  ultraviolet finite in supersymmetric vacua.

Because they characterise the effective low energy theory, the lowest dimension correlation functions  are of special interest. It is important, for instance, to understand their dependence on the different data of the string compactification. 
Of these, the two-point  function  of scalars $\phi_i$ belonging to chiral multiplets $\Phi^i$ gives information about the K\"ahler metric:
\beq
\bra \phi^{i}(k) \ov{\phi}^{i} (-k)\ket = k^2 G_{\Phi^{i},\ov{\Phi}^{i}}. 
\eeq
The computation of these two point functions is the subject of this work.

The chiral fields have a cubic superpotential $W \supset \frac{\lambda_{ijk}}{3!} \Phi_i \Phi_j \Phi_k$. In type $IIB$ string theory, $\lambda_{ijk}$ depend only on the complex structure and open string moduli, while in type $IIA$  it may only depend upon the K\"ahler and open string moduli. 
The one loop correlator can be split into field theory part and threshold correction
\beq
\Delta^{(1)} G_{\Phi^{I},\ov{\Phi}^{I}} = - \tilde{\gamma}_{\Phi_i} \log k^2/M_s^2 +  \Delta_{\Phi_I} 
\eeq
where $\tilde{\gamma}_{\Phi_i}$ is given by
\beq
\tilde{\gamma}_{\Phi_i} = \frac{|\lambda_{ijk}|^2}{16\pi^2} \frac{1}{G_{\Phi_j, \ov{\Phi}_j}G_{\Phi_k, \ov{\Phi}_k}} .
\label{deltaG}
\eeq 
and is related to the one-loop anomalous dimension $\gamma_{\Phi_i}^{(1)}$ by
\beq
\gamma_{\Phi_i}^{(1)} = \frac{\tilde{\gamma}_{\Phi_i}}{G_{\Phi_i, \ov{\Phi}_i}} = \frac{|Y_{ijk}|^2}{16\pi^2},
\eeq
where $Y_{ijk}$ is the physical Yukawa coupling.
Just as threshold corrections modify the tree-level gauge couplings, so corrections to the K\"ahler potential change the Yukawa couplings (see for example \cite{Antoniadis:1992pm}) of which the one loop corrections may be significant. Moreover, they can teach us about the expected effects of supersymmetry breaking when mediated by moduli fields.

Below, we will carry out these computations for the case of toroidal orbifold / orientifold compactification factorisable on three two tori $T^2$ of type II strings (see \cite{IIB} for reviews of IIB model building, and \cite{IIA:Reviews} for IIA, and \cite{IIA:Models} for recent work in this area) . These models provide an exceptional laboratory in this respect as they have a simple geometrical picture and they allow an explicit computation through conformal field theory techniques \cite{Friedan:1985ge, Burgess:1986ah, Garousi:1996ad}.  The chiral states are identified with the massless modes of two types of open strings: (i) open strings with both ends on the same stack of branes; and (ii) open strings with one 
end on a brane stack $a$ and the second end on a different one $b$, the two stacks intersecting at non-vanishing angles $\theta^{\kappa}_{ab}$ on 
each torus $\kappa$.
To our knowledge the one-loop two point functions have been computed only for the former case (i) in \cite{Antoniadis:1996vw, Bain:2000fb, Berg:2005ja}. This was related to the calculation of corrections to closed string moduli, as also in \cite{Antoniadis:2003sw,Bachas:1997mc,Berg:2004ek}. Moreover,  because of the diverse application of moduli fields for the determination of coupling ``constants'', supersymmetry breaking or inflationary models, previous works focused on the  contributions from the $N=2$ sectors of orbifolds where the moduli dependence appears. However the states of type (ii) localised at brane intersections play an important role, as for example,  usually the matter and  the Higgs fields in phenomenological considerations are most often  identified with them. We believe it is useful to provide the tools and expressions to compute explicitly the one-loop two point functions for these states. The computation proceeds in a different manner as it involves now computing two point functions for boundary changing operators. Using the techniques of  \cite{Dixon:1986qv} for twist operators, the computation of the tree-level  correlations of boundary changing operators have been introduced for open strings in \cite{Gava:1997jt},  successively  used,  first with four insertions with orthogonal brane intersections in \cite{Antoniadis:2000jv},  then for generic angles, the worldsheet instanton contributions was found in \cite{Cremades:2003qj} before the full CFT computation was performed in \cite{Cvetic:2003ch, Abel:2003vv, Klebanov:2003my},  and finally   generic N-point functions were calculated in \cite{Abel:2003yx}. Later, building on the general methods of \cite{Atick:1987kd}, one-loop diagrams with boundary changing operators have been constructed  \cite{Abel:2004ue,Abel:2005qn}.

Amplitudes involving boundary changing  operators are sensitive to the compactification space data (as the vacuum expectation values of the moduli) as they are suppressed by contributions of the world-sheet instantons. For tree-level amplitudes this dependence appear explicitely for three and higher points correlations. However, at higher loops, it is already present at the two-point function. Part of this dependence is due to the Yukawa couplings in equation (\ref{deltaG}), but it is interesting to investigate the remaining moduli-dependent parts.  For the case of the annulus amplitude, some steps in computing the two-point function have been  taken in this direction in \cite{Abel:2005qn}. Although the method was outlined, the computation in the case where the amplitude involved three boundaries  was not performed. Moreover, we did not find a similar computation for the M\"obius strip  available in the literature.  We will present here the relevant techniques and apply them to get explicit results.
 
The paper is organised as follows. In section 2, we will start by computing the one-loop two point function in the case of open strings with both ends on the same stack of $D$-branes when  an orbifold action leads to an $N=1$ massless theory.  This will allow us to compare the results with the case of  open strings on brane intersections. The two point function for the latter is given in the case of annulus in section 3, and in the case of Moebius strip in section 4. Some useful formulae are listed in the appendix.

\section{Warm Up: Orbifolds}

In this section, we are interested by the two point functions for the massless chiral modes of open strings propagating on  $D3$ branes  in   orbifold models. For simplicity,  the target space is   taken as $\mathbb{R}^4 \times (\mathbb{T}^2 \times \mathbb{T}^2 \times \mathbb{T}^2)/\Gamma$, where $\Gamma \subset SU(3)$ is an abelian orbifold group. The compact space is parametrised by complex pairs of coordinates $X^{I},\ov{X}^{I}$  ($I=1,2,3$) with the torus identifications:
\beq
X^{I} \sim X^{I} + 2\pi R_1^{I}, \qquad X^I \sim X^{I} + 2\pi i R_2^{I} \sin \alpha^I
\label{ORBIFOLD:torus}\eeq 
where the angle $\alpha^I$ parametrises the complex structure of the torus; alternatively the torus data is encoded in the K\"ahler and complex structure moduli respectively
\beq
T^I \equiv T_1^I + i T_2^I = iR^I_1 R^I_2 \sin \alpha^I,\qquad U^I \equiv \frac{R^I_2}{R^I_1}e^{i\alpha^I}.
\eeq
The action of an element $g$ of the orbifold  on the three compact dimensions is specified by the twist vector $(g_1,g_2,g_3) \in SU(3)$ (i.e. $\sum_{I} g_I = 0$ to preserve at least $N=1$ supersymmetry) by
\beq
g X^I = e^{2\pi i g_I} X^I.
\eeq
This projection  acts on the open strings modes leading to chiral massless states $\Phi^I$ (where $I=1,2,3$ refers to an internal dimension and group indices are suppressed)

We will not discuss the brane and orientifold content of the model which depends for instance on $N$, but  and suppose a set of   $D3$ branes can be located at the fixed points, intersecting with (a possibly vanishing number) of $D7$ branes. On each of the $D3$ brane, we assume a set of chiral states $\Phi^I$ (where $I=1,2,3$ refers to an internal dimension and group indices are suppressed).

Depending on the orbifold, there can be twist vectors $(g_1,g_2,g_3)$ lying in $SU(2)$ or $U(1)$ instead of $SU(3)$.  They lead respectively to sectors with states in $N=2$ or $N=4$ supersymmetry representations. Their contribution at one-loop to the K\"ahler potential has been calculated. For the $N=4$ case, it is found to vanish. In contrast, the $N=2$  sectors have attracted a lot of attention as they give moduli dependent results. It was computed  
in full in \cite{Berg:2005ja}. Their result for the correction to the K\"ahler metric at zero expectation value for the chiral matter fields (analagous to the one that we shall give below for $N=1$ sectors) is 
\beq
\bra \phi^{I}(k) \ov{\phi}^{I} (-k)\ket = \frac{k^2}{16\pi^2} \frac{d_{\sigma}^2}{4 T_2^I} \log (8\pi^3 \mu (k^2) T_2 U_2|\eta(U)|^4) \sum_{\sigma} \sum_g \bigg( \tr (\gamma_{\sigma,g} \lambda \lambda^{\dagger} Q_{\sigma,g}) \bigg)
\eeq
where $\mu(k^2)$ is an infra-red regulator. The parameter $\sigma$ is used to indicate the boundary conditions: $\{\sigma\} = \{\mathcal{A}^{33},\mathcal{A}^{37_I},\mathcal{M}^{33}\}$ denoting annulus diagrams between $D3-D3$, $D3-D7_I$ branes and M\"obius diagrams between $D3$ branes respectively. $d_\sigma = 1$ for annulus diagrams, $d_\sigma=2$ for M\"obius strip diagrams. Also
\beq
\gamma_{\sigma, g} \equiv \left\{ \begin{array}{cc} \gamma_{g}^3 \otimes (\gamma_g^3)^{-1} & \sigma = \mathcal{A}^{33} \\
 \gamma_{g}^3 \otimes (\gamma_g^{7_I})^{-1} & \sigma = \mathcal{A}^{37_I} \\
-(\gamma_{\Omega g}^3)^{T} (\gamma_{\Omega g}^3)^{-1}  & \sigma = \mathcal{M}^{33} \end{array} \right.
\eeq
and
\beq
Q_{\sigma, g} \equiv \prod_{J | g_J \ne 0, h_J = 0} (2\sin \pi g_J) 
\eeq
where $h_I = 0 \, \forall I$ for a $D3-D3$ partition function, and $h_I =0, h_{J\ne I} = \pm 1/2$ for $D3-D7^I$. $\sum_I h_I = 0$ is required to preserve supersymmetry.

The field theory behaviour of the $N=1$ sectors was studied on the $Z_3$ orientifold with purely $D3$ branes in \cite{Bain:2000fb}; in the following we provide a general analysis with the inclusion of the contribution of $D7$ branes, and give a closed form expression for the moduli-independent constants. We also regularise using the off-shell extension of the amplitude, which allows a direct identification of the infra-red cutoff with the physical momentum; this amplitude proves to be a good example where this technique can be easily applied, rather than, for example, zeta-function regularisation. However, it is worth mentioning that it can be shown that the same techniques apply to the calculations of \cite{Akerblom:2007np} and give the same result.

In the internal  compact space, the total  $D3$ and $D7$-brane Ramond-Ramond charges must vanish. The global cancellation  of the corresponding tadpoles reads
\beq
\sum_i tr (\gamma_{1}^{3_i}) - N_{O3}/2 = \sum_j 
\Pi_{7_j}- 8 \Pi_{O7}= 0,
\eeq
where the sum is over untwisted sectors and we have included the contribution of the orientifold planes $O3$ and $O7$;  $\Pi_A$ denotes the homology element corresponding to the cycle wrapped by the $D7$ brane or orientifold. We find, however, that the global considerations are irrelevant for the calculation of the two point corrections. More important is the twisted tadpole cancellation condition, enforced at each fixed point. 
Since 
\beq
\gamma_{\Omega g}^{-1} \gamma^{T}_{\Omega g} = \rho \gamma_{2g} 
\eeq
where $\rho = \pm 1$, we cancel M\"obius diagrams at twist $g$ with annulus diagrams at twist $2g$
\beq
\tr (\gamma_{\mathcal{A}^{33},2g} Q_{\mathcal{A}^{33},2g}) + \sum_{J} \tr (\gamma_{\mathcal{A}^{37_J},2g} Q_{\mathcal{A}^{37_J},2g}) + 4 \rho \tr (\gamma_{2g} Q_{\mathcal{M}^{33},g}) = 0 .
\eeq
This then factorises:
\beq
\tr (\gamma_{2g}^3) Q_{\mathcal{A}^{33},2g} + \sum_{J} \tr (\gamma^{7_J}_{2g}) Q_{\mathcal{A}^{37_J},2g} + 4 \rho Q_{\mathcal{M}^{33},g} = 0 .
\label{ORBIFOLD:twistRR}\eeq
Note that for certain orbifolds (such as $\mathbb{Z}_N$ with $N$ non-prime) there is a separate condition
\beq
\sum_{\sigma \ne \mathcal{M}^{33}} \tr (\gamma_{\sigma,g} Q_{\sigma,g} )= 0 \qquad | g \ne 2g'
\eeq
for elements of the orbifold not generated by $2g'$.

In the spin structures $\left[\begin{array}{c}\alpha\\ \beta \end{array}\right]$,  the partition function for annulus or M\"obius diagrams can be written (see, for example, \cite{Berg:2005ja}):
\beq
Z_g^h\left[\begin{array}{c}\alpha\\ \beta \end{array}\right] (\tau)= \frac{\eta_{\alpha\beta} }{(8\pi^2 \ap \Im(\tau))^2} \frac{1}{\eta^3 (\tau)} \tab{\alpha}{\beta} (0,\tau) \prod_{I=1}^3 Z_{int}^I\left[\begin{array}{c}\alpha\\ \beta \end{array}\right] (\tau)
\eeq
Our conventions for $\eta_{\alpha\beta}$ and  for the theta-functions given in appendix \ref{App:A} and 
\beq
Z_{int}^I \left[\begin{array}{c}\alpha\\ \beta \end{array}\right] (\tau)=  \frac{\tab{\alpha+h_I}{\beta+g_I}(0,\tau)}{\tab{1/2+h_I}{1/2+g_I}(0,\tau)} \times f(h_I)
\eeq
with
\beq
f(h_I) \equiv \left\{ \begin{array}{cc} 2\sin(\pi g_I) & h_I = 0 \\ 1 & h_I = \pm 1/2 \end{array} \right.  .
\eeq

The zero ghost picture vertex operators $V^{0}_{\Phi^I}$, $ V^{0}_{\ov{\Phi}^I}$ corresponding to the complex scalars in the chiral $\Phi^I $ and   anti-chiral
$\ov{\Phi}^I $ multiplets, respectively,  are given by
\begin{eqnarray}
V^{0}_{\Phi^I} &=& \lambda  \bigg[ 2\ap (k \cdot \psi )\Psi^{I} + \dot{X}^{I} \bigg] e^{ik \cdot X} \nonumber \\
V^{0}_{\ov{\Phi}^I} &=& \lambda^{\dagger} \bigg[ 2\ap (k \cdot \psi ) \ov{\Psi}^{I} +  \dot{\bar{X}}^{I} \bigg] e^{ik \cdot X}.
\end{eqnarray}
without any factors of the string coupling. The worldsheet fields appearing in the above are the compact coordinates $X^I (z_1), \ov{X}^I (z_2)$, their fermionic superpartners $\Psi^I (z_1) , \ov{\Psi}^I (z_2)$ and the non-compact fermionic fields $\psi^{\mu} (z_i)$.   The Chan-Paton factors $\lambda, \lambda^{\dagger}$ are determined by the orbifold projections by requiring 
\beq
\gamma_g \lambda \gamma_g^{-1} = e^{2\pi i v_g^{I}} \lambda,
\eeq

To calculate annulus and M\"obius strip diagrams, we insert the vertex operators on the imaginary axis taking $z_1=0$ and $z_2=iq$. The one-loop worldsheets  are mapped to complex plane domains defined to be $[0,1/2] \times [0,it]$ for the annulus and $[0,1/2]\times[0,2it]$ for the M\"obius strip. However, in order to sum over the diagrams, it is necessary to rescale the modular parameter for the M\"obius strip by $t \rightarrow t/4$, and so we shall in this section use $[0,1/2]\times[0,it/d_\sigma]$ as the domain. We can then write all of the diagrams in a unified way, using $\tau \equiv it$ for the annulus, and $\tau \equiv 1/2 + it/4$ for the M\"obius strip. Using the elementary Green functions 
\beq
\bra \Psi^{I} (iq) \ov{\Psi}^{I} (0) \ket^{\alpha,\beta}_{g_I,h_I} = \frac{\tab{\alpha+h_I}{\beta+g_I} (iq,\tau) \theta_1^{\prime}(0)}{\tab{1/2+h_I}{1/2+g_I} (0,\tau)\theta_1 (iq,\tau)} 
\eeq 
we obtain: 
\begin{eqnarray}
\mathcal{A}^{\sigma}_I &\equiv&  4(\ap)^2 k^2  \int_0^{\infty} dt \int_0^{t/d_{\sigma}} dq \chi(q) \nonumber \\  
&&\times \frac{\eta_{\alpha\beta} }{(8\pi^2 \ap t/d_{\sigma}^2)^2} 2\pi \frac{\theta_1^{\prime}(0)\tab{\alpha}{\beta} (iq,\tau) }{(\theta_1 (iq,\tau))^2} \frac{\tab{\alpha+h_I}{\beta+g_I}(iq,\tau)}{\tab{1/2+h_I}{1/2+g_I}(0,\tau)} f(h_I) \prod_{J \ne I} Z_{int}^J \nonumber
\end{eqnarray}
where
\beq
\chi(q) \equiv \bra e^{i k \cdot X (iq)} e^{-i k \cdot X (0)} \ket = \left(\frac{\theta_1 (iq)}{\theta_1^{\prime} (0)}e^{-\frac{d_{\sigma}^2 \pi}{t} q^2}\right)^{-2\ap k^2}.
\label{ORBIFOLD:defchi}\eeq
The two-point function of interest then reads:
\beq
\bra \Phi^I (k)\ov{\Phi}^I (-k) \ket  = \sum_{g} \mathcal{A}^{\sigma}_I \tr (\gamma_{\sigma,g} \lambda \lambda^{\dagger} Q_{\sigma, g} ) 
\eeq
Note that by using equation \ref{ORBIFOLD:twistRR} we can cancel any function that is universal to the annulus and M\"obius diagrams; we find
\beq
\tr (\gamma_{\mathcal{M}^{33},2g} \lambda \lambda^{\dagger} Q_{\mathcal{M}^{33}, g} ) \mathcal{F}(2g) + \sum_{\sigma \ne \mathcal{M}^{33}} \tr (\gamma_{\sigma,2g} \lambda \lambda^{\dagger} Q_{\sigma, 2g} ) \mathcal{F}(2g)  = 0
\eeq
for any $\mathcal{F}(2g)$.

The identity (\ref{A:thetapluscd}) and the supersymmetry conditions $\sum_I h_I = \sum_I g_I = 0$ are used to write $\mathcal{A}^{\sigma}$ as (c.f. \cite{Bain:2000fb})
\beq
\mathcal{A}^{\sigma}_I = 8\pi i (\ap)^2 \int_0^{\infty} \frac{\d t}{(8\pi^2 \ap t/d_{\sigma}^2)^2} \int_0^{t/d_\sigma} \d q e^{-2\pi h_I q} \frac{\theta_1^{\prime} (0,\tau) \theta_1 (iq + h_I \tau + g_I,\tau)}{\theta_1 (iq,\tau) \theta_1 (h_I \tau + g_I,\tau)} \chi(iq)
\eeq
which, in the closed string channel, i.e. expressed  in terms of $l = 1/t$ 
takes the form:
\begin{eqnarray}
\mathcal{A}^{3s}_I &=& 
  \int_0^{\infty} \frac{i \d l}{8\pi^3} \int_0^{1} \d x \, i\, e^{-2\pi i g_I x}\frac{\theta_1^{\prime} (0,il) \theta_1 (x + h_I - g_I il,il)}{\theta_1 (x,il) \theta_1 (h_I - g_I il ,il)} \chi(ix/l)  \\
\mathcal{M}^{33}_I &=& d_{\sigma}^2 \int_0^{\infty} \frac{i \d l}{8\pi^3} \int_0^{1} \d x \, i\, e^{-4\pi i g_I x}\frac{\theta_1^{\prime} (0,il - 1/2) \theta_1 (x  - 2 g_I il,il-1/2)}{\theta_1 (x,il-1/2) \theta_1 (- 2g_I il ,il-1/2)} \chi(ix/l) \nonumber
\end{eqnarray}
where $s = \{3,7_J\}$.
The expansion
\beq
\frac{\theta_1^{\prime} (0) \theta_1 (a+b)}{\theta_1 (a) \theta_1 (b)} = \pi \cot (\pi a) + \pi \cot (\pi b) + 4\pi \sum_{m, n = 1}^{\infty} e^{2\pi m n i \tau} \sin (2\pi m a + 2 \pi n b )
\eeq
allows the identification of two sources of infrared divergences.
The first in the open string channel, proportional to $\log k^2$, corresponds to the usual beta-function running. The second in the closed string channel, ultra-violet in the open string one,  which instead appear as a  $1/k^2$ pole preceding a divergent integral, indicating a fundamental inconsistency of the theory arising from uncancelled $RR$ charges. 
If we expand in the closed string channel, then the UV divergence can be simply subtracted; it comes entirely from the closed string zero mode. However, in this channel regulating the infra-red divergence is more subtle. Consider the behaviour of the momentum dependent part in the two regimes
\beq
\chi \rightarrow \left\{ \begin{array}{cc} (d_\sigma l)^{2\ap k^2} (2 \sin \pi x)^{-2\ap k^2} & l \rightarrow \infty \\ e^{-\frac{2\pi\ap t k^2 x(1-x)}{d_\sigma}} & t \rightarrow \infty \end{array} \right.
\label{ORBIFOLD:chilims}
\eeq
Since we are interested in the divergent and finite terms, but not those $O(k^2)$, we can split the integral into two regions, $l$ greater or less than $C \epsilon$, where $\epsilon \equiv 2\pi\ap k^2$, and $C$ is some constant.  Employing
\beq
\coth (x) = \mathrm{sign} (x) [1+ 2 \sum_{n=1}^{\infty} \exp (-2|x|n)]
\eeq
we find as $\epsilon \rightarrow 0$
\begin{eqnarray}
-8\pi^3 \mathcal{A}^{33}_I  &=& -i\int_0^1 \d x \int_0^{C \epsilon} \frac{\d l}{l} \  \pi (-i + \cot \pi g) e^{-\epsilon x(1-x)/l} \nonumber\\
&& + \int_0^1 \d x \int_{C\epsilon}^{\infty} \d l e^{-2\pi i g_I x}  4\pi \sum_{m, n = 1}^{\infty} e^{-2\pi m n l} \sin (2\pi m x - 2 \pi n g i l ) \nonumber \\
&& + \int_0^1 \d x \int_{C\epsilon}^{\infty} \d l e^{-2\pi i g_I x} \pi (\cot \pi g_I il +i) ,
\end{eqnarray}
where we have subtracted the zero mode terms without affecting the finite part of the amplitude. 

After some algebra,  taking $\epsilon \rightarrow 0$ and $C \rightarrow \infty$ such that $C \epsilon \rightarrow 0$, leads to
\begin{equation}
\mathcal{A}^{33}_I =  \frac{i}{8\pi^2} \frac{e^{-\pi i g_I}}{\sin \pi g_I} \bigg[  \log 2 \pi \epsilon - 2 + \gamma_E \bigg] \nonumber \\
 +  \frac{i}{8\pi^4} e^{-\pi i g_I} \sin \pi g_I \bigg[ \zeta' (2,1-g_I) + \zeta' (2,g_I) \bigg].
\label{ppp}
\end{equation}
In the above, the derivatives of the Hurwitz zeta function are on the \emph{first} argument, so that
\beq
\zeta'(s,a) \equiv \sum_{m=0}^{\infty} \frac{\log |m + a|}{ |m+a|^s} .
\eeq

The contribution from $D3-D7_I$ states is identical to the above. However, for $D3-D7_J$ with  $J \ne I$ we have $h_I = 1/2$ whose  contribution can be seen to be infra-red finite. Hence we can expand in the closed string channel and set $k^2=0$ directly. Expanding
\begin{eqnarray}
8\pi^3 \mathcal{A}^{37_J}_I \! \! &=& \! \! \int_0^1 \! \!\d x \int_0^{\infty} dl \pi e^{-2\pi i g_I x}\nonumber \\ & \bigg[& \cot \pi x + i \tanh \pi g l 
+  \! \!4  \! \!\sum_{m,n =1}  \! \!(-1)^n e^{-2\pi m n l} \sin (2\pi m x - 2\pi n g i l) \bigg], \nonumber
\end{eqnarray}
then subtracting the pole parts and integrating we obtain
\beq
\mathcal{A}^{37_J}_I = \frac{i}{8\pi^2} \frac{e^{-i \pi g_I}}{\sin \pi g_I} \log 2
\eeq

The contribution from M\"obius amplitudes does contain an infra-red portion; we obtain
\begin{eqnarray}
-8\pi^3 \mathcal{M}^{33}_I &=& -i\int_0^1 \d x \int_0^{C \epsilon} \frac{\d l}{l} \  \pi (-i + \cot \pi g) e^{-\epsilon x(1-x)/2l} \\
&&+  \int_0^1 \d x \int_{C\epsilon}^{\infty}\d l \pi e^{-4\pi i g_I x} \bigg[ \cot \pi x \nonumber \\
&&+ i \coth \pi 2 g_I l  + 4 \sum_{m,n=1} (-1)^{mn} e^{-2\pi m n l} \sin (2 \pi m x - 4 \pi n g_I i l ) \bigg] \nonumber 
\end{eqnarray}
which then becomes
\begin{eqnarray}
 \mathcal{M}^{33}_I &=& \frac{i}{8\pi^2} \frac{e^{-\pi i g_I}}{\sin \pi g_I} \bigg[ \log 2\pi \epsilon - 2 + \gamma_E \bigg] \nonumber \\
&& +\frac{i}{32\pi^4} \sin 2 \pi g_I e^{-2\pi i g_I} \bigg[ \zeta^{\prime} (2, 1-|g_I|) + \zeta' (2,|g_I|) + \frac{\log 2}{\cos^2 \pi g_I} \bigg]
\end{eqnarray}

We have thus computed all of the contributions to the one-loop K\"ahler metric for the states $\Phi^I$ on such orientifolds. These can be split into beta-function and threshold contributions according to the choice of renormalisation scheme that one wishes to match in the field theory. However, note that, since $M_s^{-2} = 4\pi^2 \ap$, we can rewrite in each contributino $\log 2\pi \epsilon = \log k^2/M_s^2$. It was shown in \cite{Bain:2000fb} for the $Z_3$ orientifold that the field theory result was reproduced; here we have generalised the approach slightly, included the contribution from $D7$-branes (which do not contribute to the field theory running, only the K\"ahler metric corrections) and computed the numerical corrections. It is hoped that these may have useful applications as they have a certain universal quality: since they do not depend upon the moduli, we expect them to be unaffected by implanting the singularity in a different geometry.


At the end of this section, we would like to notice that exchanging the internal direction $I$  with one of the non-compact directions takes us to the  two-point function for gauge bosons which allows to compute gauge thresholds corrections contribution from $N=1$ sector. The result is moduli independent\footnote{The moduli-dependent parts which arises from $N=2$ sectors, have been explicitely  computed for instance in  \cite{Bachas:1996zt, Antoniadis:1999ge, Lust:2003ky, Akerblom:2007np, Bianchi:2005sa}.} and is found to be:
\beq
-8\pi^3 \mathcal{A}_{GT} = 
\int_0^1 \d x \int_{0}^\infty \frac{\d t}{t} \sum_{I} \frac{\theta_1^{\prime} (h_I it + g_I)}{\theta_1 (h_I i t + g_I)} \bigg( \frac{\theta_1 (ixt,it)}{\theta_1^{\prime} (0,it)} e^{-\pi x^2 t}\bigg)^{-2\ap k^2}
\eeq
which becomes for the case of $D3$-branes:
\begin{eqnarray}
-i8\pi^2 \mathcal{A}_{GT}^{33} &=& \cot \pi g_I (2 - \gamma_E + \log C) -i \bigg[ -2i \mathrm{sign}(g_I) \frac{\log 2\pi |g_I| C \epsilon}{2\pi |g_I|}  \nonumber\\
&& -2i \sum_n \frac{\log 2\pi (n+g_I) C \epsilon}{2\pi (n+g_I)} - \frac{\log 2\pi (n-g_I) C \epsilon}{2\pi (n-g_I)} \bigg] 
\end{eqnarray}
Noting the identity
\beq
\sum_{n=1}^\infty \frac{1}{n^2 - a^2} = \frac{1-\pi a \cot \pi a}{2a^2}
\eeq
we observe that the $C$-dependent parts cancel, and we obtain
\begin{eqnarray}
\mathcal{A}_{GT}^{33} = \frac{i}{8\pi^2} \cot \pi g_I (2 - \gamma_E -\log 2 \pi \epsilon) + \frac{2g_I}{\pi} \bigg[ \zeta^{\prime} (1,|g_I|) - \zeta^{\prime} (1,1-|g_I|) \bigg] \nonumber 
\end{eqnarray}

\section{Annulus Diagrams in IIA}

In this section we compute the related amplitudes to the previous section but in type IIA string backgrounds. Here we take $D6$-branes intersecting at angles
 $\pi \theta^{\ka}_{ab}$ in the torus $\ka$ with $\ka=1,2,3$, which are the analogues of branes at blown-up orbifolds.

  In the orientifold model considered here, there are many one-loop diagrams that could conribute. They can be graphically visualised as cylinders with two boundaries: the first fixed to some brane $a$, and the second to either brane $a$ or another brane. We place the vertex operators for our chiral states both on one boundary (the amplitude vanishes if they are on opposite boundaries) just as in the previous section, but now the vertex operators differ due to the presence of boundary changing operators. We shall suppose that our chiral states are trapped at the intersection $ab$ with angles $\pi \th$. There are then three classes of two-point diagrams that can be constructed, which correspond to the three types of partition function that are possible: 
\begin{enumerate}
\item Annulus diagrams with the second boundary on brane $a$ or $b$. 
\item Annulus diagrams with the second boundary on a third brane $c$ not parallel to $a$ or $b$.
\item M\"obius strip diagrams; since there is only one boundary, there is an insertion of an orientifold operator $\Omega R$ which changes the boundary from brane $a$ to its orientifold image $a'$ (or $b$ to $b'$). 
\end{enumerate}
The diagrams of type 1 were calculated in \cite{Abel:2005qn}, where it was found that there were poles corresponding to $RR$ tadpoles just as in the orbifold case; these must cancel against similar poles in the diagrams of type 2 and 3, as we shall show. The techniques required to perform the calculation in this section - the diagrams of type 2 - were also developed there for general $N$-point correlators, but an analysis of the two-point function was lacking and is provided here. In the next section we shall compute the third type of diagram.

\subsection{Correlators of Boundary-Changing Operators}

The most non-trivial part of the calculation is that involving the boundary-changing operators; these are operators inserted into the worldsheet at a boundary that interpolate between D-branes. To understand their appearance, consider that the target space fields obey Dirichlet boundary conditions perpendicular to the branes, but Neumann along them, and once we have applied the doubling trick we have a periodic boundary condition very much like for orbifolds. On the infinite strip $[-i\infty,i\infty]\times[0,1/2]$ we extend to $[-i\infty,i\infty]\times[-1/2,1/2]$ by
\begin{equation}
\partial X (w) = \left\{ \begin{array}{cl} \partial X (w) & \Re(w) > 0 \\ -\bar{\partial} \bar{X} (-w) & \Re(w) < 0 \end{array} \right.
\end{equation}
to obtain 
\begin{equation}
\partial X (w+1) = e^{2\pi i \theta} \partial X (w).
\end{equation}
This global periodicity on the strip for an intersecting state is then mapped to a local periodicity on a worldsheet for fields in the presence of a boundary-changing operator, which represents the bosonic ground state:
\begin{eqnarray}
\partial X (w) \sigma_{\theta} (z) &\sim& (w-z)^{\theta-1} \tau_{\theta} (z) \nonumber \\
\partial \bar{X} (w) \sigma_{\theta} (z) &\sim& (w-z)^{-\theta} \tau_{\theta}^{\prime} (z).
\end{eqnarray}
They are primary operators in the theory with conformal weight $\frac{\theta}{2} (1-\theta)$.

Using these boundary changing operators we form vertex operators for the massless scalars at the intersection between branes $a$ and $b$, which we shall denote $C_{ab}$, in the $-1$ ghost picture (with $e^{-\phi}$ the bosonised ghost operators) as
\begin{eqnarray}
V^{-1}_{C_{ab}} (z_1) &=& \sqrt{2\ap G_{C_{ab},\ov{C}_{ab}}} 
e^{-\phi} e^{ik\cdot X} \prod_{\ka=1}^3 e^{i(\theta^{\ka}_{ab} - 1) H^{\ka}} \sigma_{\theta^{\ka}_{ab}} \nonumber \\
V^{-1}_{\bar{C}_{ab}} (z_2) &=& \sqrt{2\ap G_{C_{ab},\ov{C}_{ab}}} 
e^{-\phi} e^{-ik\cdot X} \prod_{\ka=1}^3 e^{-i(\theta^{\ka}_{ab} - 1) H^{\ka}} \sigma_{1-\theta^{\ka}_{ab}}
\label{ANNULUS:Vertices}\end{eqnarray}
where the intersection is specified by three angles $\th$, $\kappa=1,2,3$, and where to preserve supersymmetry $\sum_{\kappa=1}^3 \th = 0 \mod 2$.  The K\"ahler metric for these models is \cite{Lust:2004cx,Bertolini:2005qh,Akerblom:2007np,Blumenhagen:2007ip,Russo:2007tc}:
\beq
G_{C_{ab},\ov{C}_{ab}} = \bigg[\prod_{\kappa=1}^3 \left(\frac{\Gamma (|\th|)}{\Gamma(1-|\th|)}\right)^{\mathrm{sign}(\th)/2}\bigg]^{\frac{1}{\sum_{\kappa=1}^3 \mathrm{sign}(\th)}} .
\eeq
In the following we shall assume $\th \ge 0$ and thus $\sum_{\kappa=1}^3 \th = 2$, and to perform the below calculations with negative angles we can take $\th \rightarrow 1 + \th$. This is a requirement of the formalism rather than merely a choice of convenience.

To calculate the diagrams of type (2) above we must calculate the correlator on an annulus of two boundary changing operators $\sigma_{\th}, \sigma_{1-\th}$ fixed to one boundary of the worldsheet. In the target space this boundary is attached to branes $a$ and $b$, interpolating between them by absorption of an open string state. The other worldsheet boundary is fixed in the target space to a brane $c$ not parallel to $a$ or $b$ (in the parallel case the calculation is that of \cite{Abel:2005qn}). We take brane $c$ to lie at an angle $\ph$ to brane $a$ in each torus, (where to preserve supersymmetry we take $\sum_{\kappa=1}^3 \ph = 2$, although the techniques below apply also for summing to zero mod 2), and the result is a worldsheet periodicity on the annulus (taken to be $[0,1/2]\times[0,it]$ doubled to $[-1/2,1/2]\times[0,it]$ as above on the infinite strip) of
\beq
\partial X (w+1) = e^{2\pi i \ph} \partial X (w).
\eeq

Correlators are split into quantum and classical parts. The correlator
\beq
\bra \sigma_{\th} (z_1) \sigma_{1-\th} (z_2) \ket =  N(t)  (i|| W^{\ka}||)^{-1/2} \\
e^{2\pi i \ph P^{\ka}} \left(\frac{\theta_1 (z_1 - z_2)}{\theta_1^{\prime} (0)}\right)^{-\th (1-\th)} e^{-S_{cl}}
\eeq
is determined by the following quantities: 
\beq
P^{\ka} \equiv \sum_i (1/2 - \theta^i ) z_i = (1/2 - \th ) qi,
\eeq
where we have placed 
\beq
z_1 = 1/2 + iq, \qquad z_2 = 1/2,
\eeq
and
\beq
|| W^{\ka}|| = A_1^{\ka} B_2^{\ka} + A_2^{\ka} B_1^{\ka}
\eeq
where
\begin{eqnarray}
A_i &\equiv& \int_{it}^0 dz \omega_i (z) \nonumber \\
B_i &\equiv& \int_{-1/2}^{1/2} dz \omega_i (z)
\end{eqnarray}
in addition to
\begin{eqnarray}
\omega_1^{\ka} (z) &=& e^{2\pi i \ph z} \frac{\theta_2 ( z- (1-\th) q i + \ph i t)}{\theta_2 (z-iq)} \left(\frac{\theta_2 (z-iq)}{\theta_2 (z)}\right)^{\th} \nonumber \\
\omega_2^{\ka} (z) &=& e^{-2\pi i \ph z} \frac{\theta_2 ( z -\th q i - \ph i t)}{\theta_2 (z-iq)} \left(\frac{\theta_2 (z-iq)}{\theta_2 (z)}\right)^{1-\th}.
\end{eqnarray}

We also require these for the classical action, which is given by
\beq
S^{\ka}\! = \!\frac{i}{4\pi \ap} \left[ \frac{M^{AA}_{\ka} (A_1^{\ka} B_2^{\ka} \!- \!A_2^{\ka} B_1^{\ka}) - 2 B_1^{\ka} B_2^{\ka}}{(A_1^{\ka} B_2^{\ka}\! + \!A_2^{\ka} B_1^{\ka})} (v_A^{\ka})^2 + \frac{2A_1^{\ka} A_2^{\ka}}{A_1^{\ka} B_2^{\ka} + A_2^{\ka} B_1^{\ka}} (M^{AA}v_A^{\ka} - v_B^{\ka})v_B^{\ka}\right]
\label{ANNULUS:action}\eeq
where
\beq
M^{AA}_{\ka} = -2i \frac{\sin \pi \ph \sin \pi(\th+\ph)}{\sin \pi \th} = 2i \frac{\sin \pi \ph \sin \pi(2 - \th - \ph)}{\sin \pi \th} \equiv 2i F(\th,\ph) 
\eeq
and
\begin{eqnarray}
v_A^{\ka} &=& -\frac{1}{\sqrt{2}} [ n_A^{\ka} L_c^{\ka} ] \nonumber \\
v_B^{\ka} &=& i\sqrt{2} [ h^{\ka} + n_B^{\ka} \frac{4\pi^2 T_2^{\ka}}{L_c^{\ka}} ] = i\sqrt{2} F(\th,\ph)   [  b^{\ka} + n_B^{\ka} \frac{4\pi^2 T_2^{\ka}}{ F(\th,\ph)  L_c^{\ka}}] .
\end{eqnarray}
where $h^{\ka},b^{\ka}$ are the height and base of the smallest triangle $abc$, $L_c^{\ka}$ is the wrapping length of brane $c$, and $h^{\ka} = F(\th,\ph) b$.
The classical contribution is then 
\beq
e^{-S_{cl}} \equiv \sum_{n_A^{\ka}, n_B^{\ka}} e^{-S^{\ka}}
\eeq

\begin{figure}
\begin{center}
\epsfig{file=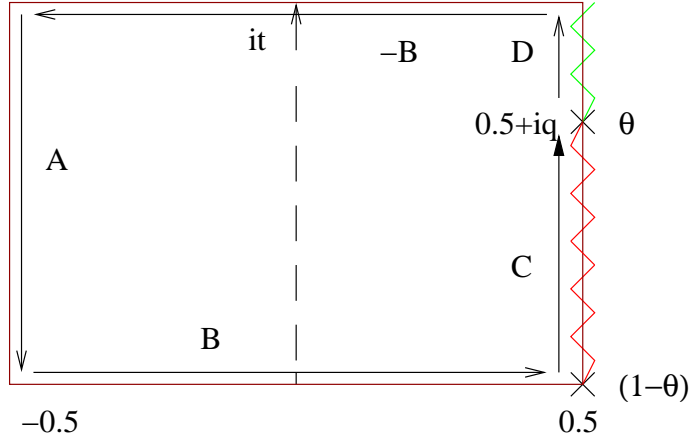}
\caption{Canonical dissection of torus.}
\end{center}
\label{2ptannulus}
\end{figure}

Note that we have written $N(t)$ for the normalisation. This shall be determined by considering the factorisation on the partition function. In doing this and in the following, we note that the integrals $A_i, B_i$ control much of the information about the amplitude, and we can use the following to help determine the $A_i$:
\begin{eqnarray}
A_{i}^{\ka} &=& -\frac{\sin \pi \th}{\sin \pi (\th + \ph)}e^{-i \pi \ph} \int_{z_1}^{it+1/2} \omega_{i}^{\ka} \equiv \frac{\sin \pi \th}{\sin \pi \ps}e^{-i \pi \ph} D_{i} \nonumber \\
&=& \frac{\sin \pi \th}{\sin \pi \ph}e^{i \pi \ps} \int_{1/2}^{z_1} \omega_{i}^{\ka} \equiv  \frac{\sin \pi \th}{\sin \pi \ph}e^{i \pi \ps} C_{i}.
\end{eqnarray}
(see fig. \ref{2ptannulus} ).

\subsubsection{Normalisation}

To normalise, we consider $q\rightarrow 0$. Note that using the above and the integral over the $C$ contour for the $A$ integrals (and the property of the theta-functions that $\theta_1 (x) = x \theta_1^{\prime}(0) + \frac{x^3}{3!} \theta_1^{\prime \prime \prime}(0) + ...$) we determine
\begin{eqnarray}
A_{1}^{\ka} &\rightarrow& \frac{\theta_1 (\ph it)}{\theta_1^{\prime} (0)} \frac{\pi}{\sin \pi \ph} \nonumber \\
A_{2}^{\ka} &\rightarrow& A_{1}^{\ka} \nonumber \\
B_{1}^{\ka} &\rightarrow& \infty \nonumber \\
B_{2}^{\ka} &\rightarrow& B_{1}^{\ka} 
\end{eqnarray}
and thus
\beq
||W|| \rightarrow 2 B_1^{\ka} \frac{\theta_1 (\ph it)}{\theta_1^{\prime} (0)} \frac{\pi}{\sin \pi \ph}.
\eeq

The classical action reduces to
\beq
S \rightarrow \frac{i}{4\pi\ap} \bigg[ -\frac{B}{A} v_A^2 + \frac{A}{B} (M^{AA} v_A - v_B) v_B \bigg].
\eeq
where we have omitted the subscripts since they become redundant in this limit. We also have $||W|| \rightarrow 2AB$. Note that we have to Poisson-resum on $n_B$ since the second term above vanishes, giving a divergent contribution after summing over $n_B$. The coefficient of $n_B^2$ is then $\frac{ -iA(4\pi^2 T_2^{\ka})^2}{B 2\pi \ap (L_c^{\ka})^2}$, and so the classical part of the boundary changing operator amplitude, plus the determinant factor, becomes
\beq
||W||^{-1/2} e^{-S_{cl}} \rightarrow \frac{e^{\pi i/4}}{A} \sqrt{\ap} \frac{L_c^{\ka}}{4\pi T_2^{\ka}} = \sqrt{\ap} \frac{L_c^{\ka}}{T_2^{\ka}} \frac{\sin \pi \ph}{4\pi^2} \frac{\theta_{1}^{\prime} (0)}{\theta_1 (\ph i t)}.
\label{ANNULUS:factorise}\eeq
Note that it can be shown that there is no zero mode contribution to the action ($n_A = n_B = 0$) as required; this can be used to show that there can be no zero mode contribution to $v_A$. We must compare this with the partition function \emph{and the OPE coefficient} $C^{(aba)}_{\th,1-\th}$. 

First consider the disk normalisation
\beq
\bra 1 \ket_a = \frac{1}{(2\pi\ap)^2} g_a^{-2}
\eeq
where $g_a$ is the Yang-Mills coupling on the brane, given by
\beq
g_a^{-2} = \frac{1}{2\pi g_s} \frac{V_a}{l_s^{p-3}}
\eeq
where $l_s = 2\pi \sqrt{\ap}$ and $V_a$ is the compact volume of the $p$-brane $a$. We require an expression for a given, internal, complex dimension, and can therefore write
\beq
\bra 1 \ket_a = \bra 1\ket_4 \prod_{\kappa=1}^3 \bra 1 \ket_\kappa
\eeq
where for $D6$-branes
\begin{eqnarray}
\bra 1\ket_4 &=& \frac{1}{(2\pi\ap)^2} \frac{1}{2\pi g_s} \nonumber \\
\bra 1 \ket_\kappa &=& \frac{L_a^\kappa}{2\pi \sqrt{\ap}}.
\end{eqnarray}
Here $L_a^{\kappa}$ is the length of the brane wrapping a one-cycle in complex dimension $\kappa$. Then since we have the freedom to normalise the wavefunctions of the vertex operators, we can use
\beq
\bra \sigma_{\th}^{ab} \sigma_{1-\th}^{ba} \ket_a = 1 = C^{(aba)}_{\th,1-\th} \bra 1 \ket_a
\eeq
to determine
\beq
C^{(aba)}_{\th,1-\th} = \frac{2\pi \sqrt{\ap}}{L_a^i}
\label{ANNULUS:OPE}\eeq

Now we wish to normalise the boundary changing operator amplitudes at one loop, so we consider
\beq
\bra \sigma_{\th}^{ab} (z_1) \sigma_{1-\th}^{ba}(z_2) \ket_{ac} \sim (z_1 - z_2)^{-\th (1-\th)} C^{(aba)}_{\th,1-\th} Z_{ac}^{X^{\ka}}
\eeq
where
\beq
Z_{ac}^{X^{\ka}} = -i I_{ac}^{\ka} \frac{\exp(\pi (\ph)^2 t) \eta(it)}{\theta_1 ( i \ph t)} .
\eeq 
Here, $I_{ac}^{\ka}$ is the number of intersections between branes $a$ and $c$ in the torus $\kappa$. Then, with the aid of the identity (\cite{Lust:2003ky})
\beq
\sin \pi \ph =  \frac{4\pi^2 T_2^{\ka} I_{ac}^{\ka} }{L_c^{\ka}L_a^{\ka}} 
\label{ANNULUS:LcLa}\eeq
we can write

\beq
N(t) = \frac{e^{\pi (\ph)^2 t}}{\eta^2 (it)}
\eeq

\subsubsection{Field Theory Limit}
\label{ANNULUS:FTL}

The field theory limit of the above amplitude is found by considering $t\rightarrow \infty$. In this regime we may expand theta-functions as 
\begin{eqnarray}
\theta_1 (z) &\rightarrow& i e^{-\pi t/4} (e^{-\pi i z} - e^{\pi i z} - e^{-2\pi t} (e^{-3 \pi i z} - e^{3 \pi i z})) + O(e^{-\pi t}) \nonumber \\
\theta_2 (z) &\rightarrow& e^{-\pi t/4} (e^{-\pi i z} + e^{\pi i z} + e^{-2\pi t} (e^{-3 \pi i z} + e^{3 \pi i z})) + O(e^{-\pi t})
\end{eqnarray}
We neglect terms $O(e^{-\pi t})$ and $O(e^{-2\pi q})$ (although retain fractional powers). We use this to determine the integrals $A_i$ and $B_i$, with the aid of the following:
\beq
\int_0^q \! dy e^{2\pi \alpha y} (1-e^{-2\pi y})^{\beta} (1-e^{-2\pi q} e^{2\pi y})^{\gamma} = \frac{e^{2\pi \alpha q}}{2\pi} B(\alpha, 1+\gamma)+ \frac{1}{2\pi} B(-\alpha,1+\beta) + O(e^{-2\pi q})
\eeq
and
\beq
\int_{-1/2}^{1/2} dx e^{\pi i \alpha x} (2\cos \pi x)^{\beta} = \frac{1}{(1+ \beta)} \frac{1}{B(1 + \frac{\beta-\alpha}{2}, 1+ \frac{\beta+ \alpha}{2})} = \frac{\Gamma(1+ \beta)}{\Gamma(1 + \frac{\beta-\alpha}{2}) \Gamma (1+ \frac{\beta+ \alpha}{2})} .
\eeq
This allows us to determine (for $\ph > 0$)
\begin{eqnarray}
A_{1}^{\ka}\!\! \!&\rightarrow& \!\!\!\! -\frac{i}{2\pi} e^{\pi \ph t} \bigg[ e^{-2\pi (1-\th) q } B(\th\! + \!\ph \!-1, 1-\!\ph) + e^{-2\pi \ph q } B(1\!-\!\th \!- \!\ph,\ph) \bigg]\nonumber \\
&& \nonumber \\
A_{2}^{\ka}\!\! \!&\rightarrow& \!\!\!\! -\frac{i}{2\pi} e^{\pi \ph t}\bigg[  e^{2 \pi q (\th + \ph - 1)} B (\ph + \th - 1,1-\ph) + B(1-\th-\ph,\ph) \bigg]\nonumber \\
&&   \nonumber \\
B_1^{\ka} \!\! \!&\rightarrow& \!\!\!\!e^{-2\pi q (1-\th)  +\pi \ph t } \frac{\Gamma (1-\th)}{\Gamma (\ph) \Gamma (2- \th - \ph)} + e^{-\pi \ph t} \frac{\Gamma(1-\th)}{\Gamma (1+ \ph) \Gamma (1 - \th - \ph)} \nonumber \\
B_2^{\ka} \!\! \!&\rightarrow& \!\!\!\!e^{\pi \ph t} \frac{\sin \pi \phi}{\pi} \bigg[ B(\ph, \th) - e^{2\pi (q\th+(\ph-1)t)} B(\ph-1,\th)\bigg] \nonumber \\
&\rightarrow& \!\!\!\! \frac{e^{\pi \ph t}}{B(1-\ps,1-\ph)} \bigg[ \frac{1}{1-\ps} + \frac{e^{2\pi (q\th+(\ph-1)t)}}{1-\ph} \bigg]
\label{PrecisionABs}\end{eqnarray}

The leading behaviour of $||W||$ (where for simplicity in the following we shall take $\th, \ph > 1/2$) is given by
\beq
||W|| \rightarrow  -\frac{i}{2\pi} \exp [2 \pi \ph t] \exp[ 2\pi q (\th - \ps) ] \Gamma^{\ka} 
\eeq
where we have defined
\beq
\Gamma^{\ka} \equiv \frac{ \Gamma (1-\th)\Gamma (1-\ph) \Gamma (1-\ps)}{\Gamma (\th) \Gamma (\ph)\Gamma (\ps)}
\eeq
for later use; note that $G_{C_{ab},\ov{C}_{ab}}G_{C_{bc},\ov{C}_{bc}}G_{C_{ca},\ov{C}_{ca}} = \prod_{\ka} 
(\Gamma^{\ka})^{1/2}$.

In this limit, $A_2 B_1$ dominates over $A_1 B_2$ and we obtain for the classical action
\beq
S^{\ka} \rightarrow \frac{i}{4\pi \ap} \left[ -M^{AA}_{\ka} (v_A^{\ka})^2 + \frac{2A_1}{B_1} (M^{AA}_{\ka}v_A^{\ka} - v_B^{\ka})v_B^{\ka}\right] 
\eeq
 Noting that $\frac{A_1^{\ka}}{B_1^{\ka}} \rightarrow 1/M_{AA} $, we obtain
\beq
S^{\ka} \rightarrow \frac{i}{4\pi \ap} \left[ -M^{AA} \left( v_A - \frac{v_B}{M^{AA}} \right)^2 - \frac{v_B^2}{M^{AA}} \right]
\eeq
which gives us
\begin{eqnarray}
S \!\!\!\!&\rightarrow&\!\! \!\frac{1}{2\pi \ap} \frac{1}{2} F(\th,\ph)  \bigg[\! (b^{\ka} + n_A^{\ka} L_c^{\ka} + n_B^{\ka} \frac{4\pi^2 T_2^{\ka}}{ F(\th,\ph)  L_c^{\ka}})^2 \!+ \!(b^{\ka} + n_B^{\ka} \frac{4\pi^2 T_2^{\ka}}{ F(\th,\ph)  L_c^{\ka}})^2\bigg]\nonumber \\
\!\!&\rightarrow& \!\!\frac{1}{2\pi \ap} [ A(n_A,n_B) + A(n_B) ]
\end{eqnarray}
This is just two sums over areas of triangles $abc$ wrapping the torus, and gives the expected field theory factor as the product of two Yukawa couplings. Note the similarity to the tree level expression as given for instance by equation (A.17) of \cite{Abel:2005qn}.

\subsubsection{Fermionic Part}

Accompanying the bosonic amplitude is the fermionic one. The correlators are given by
\beq
\bra \prod_{j} e^{ia_i H(z_j)} \ket_{\nu} = e^{2\pi i \alpha Q} \theta_{\nu} (\alpha i t + Q) \prod_{i < j} \left(\frac{\theta_1 (z_i - z_j)}{\theta_1^{\prime} (0)}\right)^{a_i a_j}
\eeq
where
\beq
Q \equiv \sum_i a_i z_i.
\eeq
This gives for us 
\beq 
Q = (\th - 1) (z_1 - z_2) = (\th - 1) qi
\eeq for the operator  $e^{i (\th - 1)H } e^{-i (\th - 1) H}$, while for $e^{i \th H} e^{-i \th H}$ we have
\beq
Q^{\prime} = \th q i .
\eeq 
Note that the fermionic partition function is
\beq
Z^{\psi^{\ka}}_{ac} = -i\frac{\theta_{\nu} (i \ph t) }{\exp [\pi (\ph)^2 t] \eta (it)}.
\eeq
and thus we require a normalisation factor of $i \eta(it)^{-1} \exp [-\pi (\ph)^2 t]$.

\subsection{Full $N=1$ Amplitude}

We have now assembled all of the ingredients to write down the full amplitude. This is 
\begin{multline}
\mathcal{A} \equiv \bra C_{ab} (k)\bar{C}_{ab}(-k) \ket_c = G_{C_{ab},\ov{C}_{ab}} N_c tr(\lambda_{ab} \lambda_{ab}^{\dagger})  4(\ap)^2 k^2\\
\int_0^\infty \frac{dt}{(8\pi^2\ap t)^2} \frac{1}{\eta^3 (it)}\int_0^t d q \chi(qi) e^{2\pi q} \left(\frac{\theta_1(iq)}{\theta_1^{\prime}(0)}\right)^{-2}  \sum_{\nu} \delta_\nu \frac{1}{2}\bigg[ \theta_\nu (qi) + \theta_{\nu} (-qi)\bigg] \\
\prod_{\ka = 1}^3 \frac{\theta_\nu (qi(\th - 1) + \ph i t) }{ \eta^3 (it)}|W^{\ka}|^{-1/2} \sum_{n^{\ka}_A,n^{\ka}_B} e^{-S(n_A^{\ka},n_B^{\ka})}
\end{multline}
where $\chi$ is as defined in (\ref{ORBIFOLD:defchi}).

After summing over spin structures $\nu$ we find
\begin{multline}
\mathcal{A} =  \frac{ k^2}{16\pi^2} G_{C_{ab},\ov{C}_{ab}} N_c tr(\lambda_{ab} \lambda_{ab}^{\dagger}) \int_0^\infty \frac{dt}{t^2} \frac{1}{\eta(it)^{6}} \int_0^t d q e^{2\pi q} \chi(qi)  \theta_1(iq)^{-1}    \\
\prod_{\ka = 1}^3 \theta_1 (qi(\th - 1) + \ph i t) |W^{\ka}|^{-1/2} \sum_{n^{\ka}_A,n^{\ka}_B} e^{-S(n_A^{\ka},n_B^{\ka})}.
\label{annulus2pt}\end{multline}
This, with the expression (\ref{ANNULUS:action}) is the main result of this section. We see that all of the moduli dependence is contained in the classical action.

Note that $\chi \sim (qi)^{-2\ap k^2}$ as $q \rightarrow 0$ and $\chi \sim (t-q)^{-2\ap k^2}$ as $q \rightarrow t$ and thus the above amplitude has poles at $q=0,t$, as predicted in \cite{Abel:2005qn}. Using equation (\ref{ANNULUS:factorise}) we can see that the prediction there is exactly correct, and we find 
\begin{eqnarray}
\mathcal{A} =  \frac{G_{C_{ab},\ov{C}_{ab}}}{32\pi^3} \tr(\lambda_{ab} \lambda_{ab}^{\dagger})\bigg[ \frac{(2\pi \sqrt{\ap})^3}{2\ap} \bigg( \frac{N_c I_{ac}}{L_a} + \frac{N_c I_{bc}}{L_b}\bigg) \int_0^{\infty} \frac{dt}{t^2} + \mathrm{finite}\bigg]
\end{eqnarray}

\subsubsection{Field Theory Limit}

If we now wish to take the field theory limit of the expression (\ref{annulus2pt}) we must consider $t \rightarrow \infty$. Using the expressions from section \ref{ANNULUS:FTL} and equation (\ref{ORBIFOLD:chilims}), we easily derive
\begin{eqnarray}
\mathcal{A} = tr(\lambda_{ab} \lambda_{ab}^{\dagger}) \bigg\{&& G_{C_{ab},\ov{C}_{ab}}   \frac{ N_c k^2}{16\pi^2} \bigg [\prod_{\ka=1}^3 (\Gamma^{\ka})^{-1/2} |\lambda|^2 \int_{1/2\pi \ap\Lambda^2}^\infty \frac{dt}{t} e^{-2\pi \ap t k^2 x(1-x)} \bigg] \nonumber \\ &+& k^2 \Delta G_{C_{ab},\ov{C}_{ab}} \bigg\}
\end{eqnarray}
where 
\beq
|\lambda|^2 \equiv \prod_{\ka=1}^3 \sum_{n_A^{\ka},n_B^{\ka}} \sqrt{2\pi} e^{-\frac{A(n_A,n_B)}{2\pi \ap} - \frac{A(n_B)}{2\pi \ap}}
\eeq
is the square of the coupling appearing in the superpotential. $\Delta G_{C_{ab},\ov{C}_{ab}}$ is the correction to the K\"ahler metric from integrating out the massive string modes. Note that we have used a different cutoff scheme here to section 1; here we cannot claim that there is no contribution from massive modes in the region $[1/2\pi \ap \Lambda^2,\infty]$ of $t$, but instead these give finite contributions to $\Delta G_{C_{ab},\ov{C}_{ab}}$. It would be very interesting to compute this correction, but it is complicated by, among other issues, the explicit summation over worldsheet instantons. Note that the classical action is only a constant in the field theory limit; in general it is a function of the worldsheet coordinates and the modular parameter, and so should give interesting dependence on the K\"ahler moduli to the full amplitude.

Performing the integration in the above we obtain
\begin{eqnarray}
\mathcal{A} &=& -N_c tr(\lambda_{ab} \lambda_{ab}^{\dagger})\frac{ |\lambda|^2 }{16\pi^2} \frac{1}{ G_{C_{bc},\ov{C}_{bc}}G_{C_{ca},\ov{C}_{ca}}} k^2 \bigg( \log k^2/\Lambda^2 - 2 +\gamma_E  \nonumber \\
&& - \frac{k^2}{\Lambda^2} {}_2 F_2 (1,2;2,5/2;-k^2/4\Lambda^2)\bigg)+ k^2 \Delta G_{C_{ab},\ov{C}_{ab}}.
\end{eqnarray}
This reproduces exactly the field theory result for the anomalous dimension of the superfields.

\section{The M\"obius Strip Amplitude in IIA}

In this section we provide new techniques to calculate M\"obius strip amplitudes for states at intersections between branes. This further generalises the techniques that were developed for periodic closed string amplitudes in \cite{Atick:1987kd}, were first applied to the case of intersecting branes in \cite{Abel:2004ue} and we generalised to the case of generic annulus diagrams (i.e. with no restrictions upon the angles of the branes) in \cite{Abel:2005qn}. 

\subsection{Worldsheet Periodicity}

A M\"obius strip can be considered to be a strip closed under an orientation reversal: consider 
\begin{equation}
X(w+it,\bar{w}-it) = \Omega X(w,\bar{w}) = X(1/2 - \bar{w},1/2-w).
\end{equation}
If we now combine this with a reflection to make an orientifold model
\begin{equation}
X(w+it,\bar{w}-it) = \Omega R X(w,\bar{w}) = \bar{X}(1/2 - \bar{w},1/2-w).
\end{equation}
we see that we can consistently combine this with the doubling trick for intersecting brane models. We align the coordinate system along the orientifold plane, so that for worldsheet the strip $[0,it]\times[0,1/2]$, on the imaginary axis we have Neumann boundary conditions along $X + e^{2\pi i \phi_{a,O6}} \bar{X} - c ( 1 + e^{2\pi i \phi_{a O6}})$ with Dirichlet conditions perpendicular, and along the axis $\Re(w) = 1/2$ we have Neumann conditions along $X + e^{-2\pi i \phi_{a,O6}}\bar{X} - c ( 1 + e^{-2\pi i \phi_{a O6}})$, where $c$ is the position of the intersection along the $O6$ plane. We then have boundary conditions $\partial X = -e^{\pm2\pi i \phi_{a, O6}} \bar{\partial} \bar{X}, \partial \bar{X} = -e^{\mp2\pi i \phi_{a, O6}} \bar{\partial} X$ where the upper (lower) sign is for $\Re(w) = 0 (1/2)$. Using the doubling trick
\begin{equation}
\partial X (w) = \left\{ \begin{array}{cl} \partial X (w) & \Re(w) > 0 \\ -e^{2\pi i \phi_{a,O6}}\bar{\partial} \bar{X} (-\bar{w}) & \Re(w) < 0 \end{array} \right.
\end{equation}
and similarly for $\partial \bar{X}, \bar{\partial} X$, we arrive at the new periodicity conditions
\begin{eqnarray}
\partial X (w+1) &=& e^{-4\pi i \phi_{a,O6}} \partial X (w)\nonumber \\
\partial X (w+1/2+it) &=&  e^{-2\pi i \phi_{a,O6}}\partial X (w) \nonumber \\
\partial \bar{X} (w+1) &=& e^{4\pi i \phi_{a,O6}} \partial \bar{X} (w)\nonumber \\
\partial \bar{X} (w+1/2+it) &=&  e^{2\pi i \phi_{a,O6}}\partial \bar{X} (w).
\label{Periodicities}\end{eqnarray}
This provides a convenient way to obtain the holomorphic differentials with given boundary conditions. Note that these lead to 
\beq 
\partial X (w + 2it) = \partial X (w), \qquad \partial \bar{X} (w + 2 i t) = \partial \bar{X} 
\eeq
To compute the worldsheet instanton contribution, we integrate the (doubly periodic) function $\partial X\bar{\partial}\bar{X} (w, \bar{w})$ over the fundamental domain  - but it is more convenient to extend this to the domain $[0,2it]\times[0,1/2]$, and take half of the resulting action. 

If there is also an orbifold projection, we may combine the action with the orientifold as above and adapt the doubling trick accordingly, or we can simply align our coordinate system relative to the new fixed planes.

\subsection{Cut Differentials}

The cut differentials with the periodicities (\ref{Periodicities}) are given by using the theta-function
\begin{eqnarray}
\theta \left[ \begin{array}{c} 1/2 - 2a \\ 1/2 + a \end{array} \right]\! (z+m;\tau) \!\!&=& \!\!\exp (2\pi i (1/2-2a)m)\ \  \theta \left[ \begin{array}{c} 1/2 - 2a \\ 1/2 +a\end{array} \right] (z;\tau)  \\
\theta \left[ \begin{array}{c} 1/2 - 2a\\ 1/2 + a\end{array} \right] \!(z+m\tau;\tau) \!\!&=& \!\!\exp(-2\pi i a) \exp (-2\pi i m/2) \nonumber \\ && \times \exp (-\pi i m^2 \tau - 2 \pi i mz) \ \ \theta \left[ \begin{array}{c} 1/2 -2a \\ 1/2 +a \end{array} \right] (z;\tau) \nonumber.
\end{eqnarray}
We thus define
\begin{eqnarray}
\theta_{\phi_{a,O6}} (z) &\equiv& \theta \left[ \begin{array}{c} 1/2 - 2\phi_{a,O6} \\ 1/2 + \phi_{a,O6} \end{array} \right] (z;\frac{1}{2} + it) \nonumber \\
\theta_1 (z) &\equiv& \theta \left[ \begin{array}{c} 1/2 \\ 1/2 \end{array} \right] (z;\frac{1}{2} + it).
\end{eqnarray}
For a correlator of $L$ vertex operators at coordinates $z_i$ (all lying on the imaginary axis), each with angles $\theta_i$ we may take $\sum_{i=1}^L \theta_i = M$.We then have $L-M$ cut differentials as a basis for $\partial X$, with $\{i'\} = \{1,...,L-M\}$:
\begin{equation}
\tilde{\omega}_{i'} (z) = \gamma_X (z) \theta_{+\phi_{a,O6}} (z-z_{i'} - Y) \prod_{j \in \{\alpha\}\ne \alpha}^{L-M} \theta_1 (z-z_{j})
\label{TwistOmegaa}\end{equation}
and we have the set of $M$ differentials for $\partial \bar{X}$ with $\{i''\} = \{L-M+1,...,L\}$:
\begin{equation}
\tilde{\omega}_{i''} (z) = \gamma_{\bar{X}} (z) \theta_{-\phi_{a,O6}} (z-z_{i''} + Y) \prod_{j \in \{\beta\}\ne \beta}^{L} \theta_1 (z-z_{j}).
\label{TwistOmegab}\end{equation}
Here
\begin{eqnarray}
Y &=&  -\sum_{i'} \theta_{i'} \ z_{i'}  + \sum_{i''} (1-\theta_{i''} ) z_{i''} \nonumber \\
\gamma_X (z) &=& \prod_{i=1}^L \theta_1 (z-z_i)^{\theta_i - 1} \nonumber \\
\gamma_{\bar{X}} (z) &=& \prod_{i=1}^L \theta_1 (z-z_i)^{-\theta_i}.
\end{eqnarray}

These cut differentials are a natural basis which is convenient for deriving the quantum part of the amplitude, but for performing calculations it is convenient to express the above only in usual theta functions. We replace
\beq
\theta_{\pm\phi_{a,O6}} (z-z_{\alpha} \mp Y) \rightarrow \exp [ \mp 4\pi i \phi_{a,O6} z ] \theta_1 (z - z_{\alpha} \mp Y \mp 2 i \phi_{a,O6} t).
\eeq
where we use $z_\alpha$ to denote a member of $z_{i'}$ or $z_{i''}$. 
We shall denote the new basis $\{\omega_{i'},\omega_{i''}\}$. To convert between the two bases, we have
\begin{eqnarray}
\tilde{\omega}_{i'} &=& e^{-2\pi i\phi_{a,O6} (  1 + \phi_{a, O6} (1-2it))} e^{ 4 \pi i \phi_{a,O6} (z_\alpha + Y )} \omega_{i'} \nonumber \\
\tilde{\omega}_{i''} &=& e^{-2\pi i\phi_{a,O6} ( - 1 + \phi_{a, O6} (1-2it))} e^{- 4 \pi i \phi_{a,O6} (z_\alpha - Y )} \omega_{i''}
\label{convertbasesmob}\end{eqnarray}
In this basis, the cut differentials for a two-point function with vertices at $z_1 = 0, z_2 = iq$ and angles $\theta, 1-\theta$ are  
\begin{eqnarray}
\omega_1 (z) &=& e^{-4\pi i \phi_{a,O6} z} \frac{\theta_1 ( z - \theta q i - 2\phi_{a,O6} i t)}{\theta_1 (z-iq)} \left(\frac{\theta_1 (z-iq)}{\theta_1 (z)}\right)^{1-\theta} \nonumber \\
\omega_2 (z) &=& e^{4\pi i \phi_{a,06} z} \frac{\theta_1 ( z - (1-\theta) q i + 2\phi_{a,O6} i t)}{\theta_1 (z-iq)} \left(\frac{\theta_1 (z-iq)}{\theta_1 (z)}\right)^{\theta}.
\end{eqnarray}
In each of the above, the modulus of the theta functions is $\tau = 1/2 + it$.  

\subsection{Classical Solutions}

The classical solutions $X_{cl}, \bar{X_{cl}}$ satisfy the boundary conditions
\begin{eqnarray}
\int_{\gamma_a} \d z \partial X + \d \bar{z} \bar{\partial} X &=& v_a \nonumber \\
\int_{\gamma_a} \d z \partial \bar{X} + \d \bar{z} \bar{\partial} \bar{X} &=& \bar{v}_a
\end{eqnarray}
where the $v_a$ are $L$ displacements corresponding to the independent paths $\gamma_a$ on the worldsheet. We can use these to determine the classical solutions in terms of the basis of cut differentials:
\begin{eqnarray}
\partial X_{cl} (z)&=& v_a (W^{-1})^a_{i^{\prime}} \omega_{i^{\prime}} (z) \nonumber \\
\bar{\partial} X_{cl}(\bar{z}) &=& v_a (\ov{W}^{-1})^a_{i^{\prime\prime}} \bar{\omega}_{i^{\prime\prime}} (\bar{z}), 
\end{eqnarray}
where we have defined the matrix $W$ as 
\begin{eqnarray}
W^{i^{\prime}}_a &= \int_{\gamma_a} \d z \omega^{i^{\prime}} (z), \qquad &i^{\prime} = \{1..L-M\} \nonumber \\
W^{i^{\prime\prime}}_a &= \int_{\gamma_a} \d \bar{z} \bar{\omega}^{i^{\prime\prime}} (\bar{z}), \qquad &i^{\prime\prime} = \{L-M+1..L\}. 
\end{eqnarray}
We can then use these to determine $\bar{\partial} \bar{X}_{cl}, \partial \bar{X}_{cl}$ via the doubling trick:
\begin{eqnarray}
\bar{\partial} \bar{X}_{cl} (\bar{z})&=& -e^{-2\pi i \phi_{a, O6}} v_a (W^{-1})^a_{i^{\prime}} \omega_{i^{\prime}} (-z) \nonumber \\
\partial \bar{X}_{cl}(z) &=& -e^{2\pi i \phi_{a, O6}} v_a (W^{-1})^a_{i^{\prime\prime}} \bar{\omega}_{i^{\prime\prime}} (-\bar{z}). 
\end{eqnarray}
However, we may also note that the complex conjugates of the cut differentials are a good basis for $\bar{\partial} \bar{X}_{cl}, \partial \bar{X}_{cl}$ if we extend those fields to $\Re(z) = [-1/2,1/2]$ via
\begin{equation}
\bar{\partial} \bar{X} (\bar{w}) = \left\{ \begin{array}{cl} \bar{\partial} \bar{X} (\bar{w}) & \Re(\bar{w}) > 0 \\ -e^{-2\pi i \phi_{a,O6}}\partial  X (-w) & \Re(w) < 0 \end{array} \right.
\end{equation}
to obtain
\begin{eqnarray}
\bar{\partial} \bar{X}_{cl} (\bar{z})&=& \bar{v}_a (\ov{W}^{-1})^a_{i^{\prime}} \bar{\omega}_{i^{\prime}} (\bar{z}) \nonumber \\
\partial \bar{X}_{cl}(z) &=& \bar{v}_a (W^{-1})^a_{i^{\prime\prime}} \omega_{i^{\prime\prime}} (z). 
\end{eqnarray}
These are entirely consistent provided that we choose the $v_a$ correctly. This is a crucial point: we are not at liberty to choose arbitrary cycles for the $\gamma_a$, but must match them to displacements with the correct phase. We can write 
\beq
\bar{\omega}_{i'} (\bar{z}) = e^{i\xi_{i'}} \omega_{i'} (-z), \qquad \omega_{i''} (z) = e^{i\xi_{i''}} \bar{\omega}_{i''} (-\bar{z})
\eeq
and
\begin{eqnarray}
\int_{\gamma_a} \d \bar{z} \bar{\omega}_{i'} (\bar{z}) &=& - e^{i \eta_a} e^{i\xi_{i'}} \int_{\gamma_a} \d z \omega_{i'} (z) \nonumber \\
\int_{\gamma_a} \d z \omega_{i''} (z) &=& - e^{i \eta_a} e^{i\xi_{i''}} \int_{\gamma_a} \d \bar{z} \bar{\omega}_{i''} (\bar{z})
\end{eqnarray}
where $\eta_a$ is a phase, constant across $i$ for each $a$, and such that we can write
\beq
\ov{W}^{a}_i =  - e^{i \eta_a} e^{i\xi_{i}} W^a_i .
\eeq
Then
\beq
(\ov{W}^{-1})^a_{i} = - e^{-i \eta_a} e^{-i\xi_{i}} (W^{-1})^a_{i}
\eeq
and thus we require for consistency
\beq
\bar{v}_a = e^{-2\pi i \phi_{a,O6}} e^{i \eta_a} v_a . 
\eeq
Note that the phases $e^{\xi_i}$ are always removed from amplitudes (corresponding to normalisation of the basis functions), and indeed, when we take the basis $\{\omega_{i'},\omega_{i''}\}$ they are equal to $1$ anyway. However, as mentioned the phases $\eta_a$ are crucial. A consequence of the above is that
\begin{equation}
\partial X (-z) = -e^{2\pi i \phi_{a, O6}} \bar{\partial} \ov{X} (\bar{z}), 
\end{equation}
and therefore
\begin{equation}
\int_{it+1/2}^{it} \d x \ \partial X  (x) = \int_{0}^{1/2} \d x \  \bar{\partial} \ov{X} (x).
\end{equation}
Using this, we define
\beq
\int_{\gamma_B} \d z \omega_{i'} (z) \equiv \frac{1}{\cos \psi/2} \bigg[ \int_{0}^{1/2} \d z \omega_{i'} (z) + e^{ i \psi} \int_{it}^{it+1/2} \d z \omega_{i'}(z)\bigg] .
\eeq
which we denote
\beq
\gamma_B \equiv \frac{1}{\cos \psi/2} \bigg([0,1/2] + e^{ i \psi} [it,it+1/2] \bigg).
\eeq
The above then results in 
\beq
\eta_B = \pm \pi + 2 \pi \phi_{a, O6} -  \psi,
\eeq
which gives us the phase of $v_B$,
and thus
\beq
v_B = i e^{i \psi/2}  \hat{v}_B . 
\eeq
Here $\hat{v}_B $ is a real number corresponding to the distance traversed by the cycle, and thus it may be negative. 
However, we have an apparent freedom in choosing $\psi$. This freedom is fixed by the requirement that the action not depend upon the linearly independent combination
\beq
\gamma_{\tilde{B}} \equiv \frac{1}{\cos \psi/2} \bigg([0,1/2] - e^{ i \psi} [it,it+1/2] \bigg),
\eeq
as shall be seen in the next subsection.

To fix $\hat{v}_B$, however, we must consider from the above that 
\beq
v_B = \frac{e^{i\psi/2}}{\cos \psi/2} \bigg( e^{-i\psi/2} \Delta_B X - e^{i\psi/2} \Delta_B \bar{X} \bigg).
\eeq

\subsection{Classical Action}

The classical action is determined by integrating the classical solutions over the surface:
\begin{eqnarray}
S_{cl} &=& \frac{1}{4\pi\ap} \int_R \d^2 z (\partial X \bar{\partial} \bar{X} + \bar{\partial} X \partial \bar{X}) \nonumber \\
&\equiv& \frac{1}{4\pi\ap} v_a \bar{v}_b \bigg[ (W^{-1})^a_{i'} (\ov{W}^{-1})^{b}_{j'} (\omega_{i'},\omega_{j'} ) + (W^{-1})^a_{i''} (\ov{W}^{-1})^{b}_{j''} (\omega_{j''},\omega_{i''} )\bigg] \nonumber \\
&\equiv& \frac{1}{4\pi \ap} (v^{\dagger})_b S^{ba}_{cl} v_a
\end{eqnarray}
where the region $R$ is the doubled M\"obius strip  $[0,1/2] \times [0,2it]$, and we have divided by two; the functions $\partial X \bar{\partial} \bar{X}$ and $\bar{\partial} X \partial \bar{X}$ are even under $z \rightarrow -\bar{z}$ and $z \rightarrow z + 1/2 + it$. It remains to determine the inner products $(\omega_{i'},\omega_{j'})$. To do this we perform a canonical dissection by writing $\omega_i (z) = \d f_i (z)$  and integrate using Green's Theorem, as in \cite{Atick:1987kd,Abel:2005qn}. We split the worldsheet up into paths and use Cauchy's theorem to express these in terms of the same cycles $\gamma_a$. Two paths are eliminated; the most expedient to eliminate depend upon the precise configuration, and hence we shall provide the procedure and the expressions for the two point function, rather than the general case. In the two point function, we have one vertex fixed at $z=0$, and one at $z=qi$. The range of $q$ is $[0,2t]$. The appropriate contours to take depend upon whether the initial brane is parallel to the orientifold plane or intersects with it, and whether $q > t$. 

Suppose that the first $N_D$ vertex operators have $\Im(z_i) < t$, and the following $N_C$ have $\Im(z_i) > t$, ordered in increasing $\Im(z_i)$; they all lie upon the imaginary axis, and so we define the contours
\begin{eqnarray}
C_{N_C+1} &\equiv& [2it,z_{N_C+N_D}] \nonumber \\
C_{i} &\equiv& [z_{i+1},z_i], \qquad N_D + 1 < i < N_C + N_D \nonumber \\
C_{1} &\equiv& [z_{N_D+1}, it] \nonumber \\
D_{N_D+1} &\equiv& [t,z_{N_D}] \nonumber \\
D_{i} &\equiv& [z_{i+1},z_i], \qquad  i < N_D \nonumber \\
\hat{B} &\equiv& [0,1/2] \nonumber \\
\tilde{\hat{B}} &\equiv& [it,it+1/2]
\end{eqnarray}
noting that $z_1 = 0$. 

\begin{figure}
\begin{center}
\input{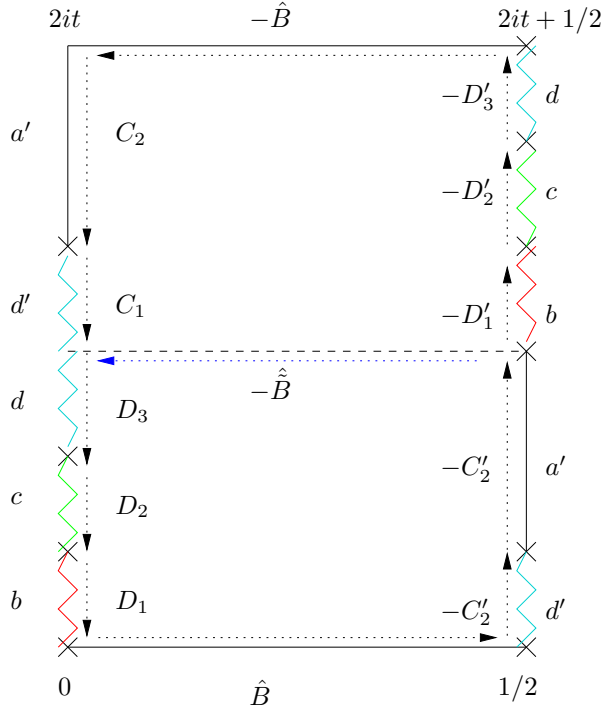}
\caption{Canonical dissection of doubled M\"obius strip.}
\end{center}
\label{MOEBIUS:dissection}
\end{figure}

We also require the conjugate contours 
\begin{eqnarray}
C_i^{\prime} &\equiv& C_i + 1/2 - it \nonumber \\
D_i^{\prime} &\equiv& D_i + 1/2 + it .
\end{eqnarray}
and the phases
\begin{eqnarray}
C_i^{\prime} &=& e^{-2\pi i \phi_{a, O6}} e^{ic_i} C_i \nonumber \\
D_i^{\prime} &=& e^{-2\pi i \phi_{a, O6}} e^{id_i} D_i
\end{eqnarray}
where
\begin{eqnarray}
d_i - d_{i-1} =& - 2\pi \theta_i \qquad &\mod 2\pi \nonumber \\
c_i - c_{i-1} =& - 2\pi \theta_{i+ N_D} \qquad &\mod 2\pi
\end{eqnarray}
and similarly for $d_i$. We have $ c_{N_C + 1} = 0$ and thus
\begin{eqnarray}
d_i &=& -2\pi \sum_{j=1}^i \theta_j \nonumber \\
c_i &=& 2\pi \sum_{j=i+1}^{N_C+i+1} \theta_{j+N_D} 
\end{eqnarray}
The configuration is illustrated in figure \ref{MOEBIUS:dissection}.

The conditions for eliminating paths are
\begin{eqnarray}
\sum_{i} C_i + \sum_j D_j &=& \sum_{i} C_i^{\prime} + \sum_j D_j^{\prime} \label{CCPrimeSum} \\
\sum_{i} C_i - \sum_j D_j^{\prime} &=& \hat{B} - \tilde{\hat{B}} 
\end{eqnarray}

Once all spurious degrees of freedom have been eliminated, we can finally write
\begin{eqnarray}
(\omega_{i'},\omega_{j'}) &=& i W^{i'}_a \ov{W}^{j'}_b M^{ab} \nonumber \\
(\omega_{i''},\omega_{j''}) &=& i \ov{W}^{i''}_a W^{j''}_b \ov{M}^{ab}
\end{eqnarray}
where $M^{ab}$ is anti-hermitian. We can  simplify by using the matrix
\beq
P_{ij} \equiv \left\{ \begin{array}{cc} \delta_{i'j'} & i \in \{i'\} \\ 0 & i \in \{i''\} \end{array} \right.
\eeq 
and defining $\hat{W}_{a i} $ by
\beq
W^a_i = i e^{- i \eta_a/2} \hat{W}_{a i} 
\eeq
(which factors out the phases for the individual cut differentials; as we argued they disappear from the action anyway - although note that it does not exclude elements $\hat{W}_{a i }$ from being negative) to then write the action in matrix form as
\beq
S^{ab}_{cl} = i [ M^T \Wh P \Wh^{-1} - (M^T \Wh P \Wh^{-1})^{\dagger} - M^T   ]^{ab}
\eeq

\subsection{Classical Action: Two-Point case}

It is possible to deal with the two point functions quite generally; initially we have five paths $C_i, D_i, \hat{B}, \tilde{\hat{B}}$ where $N_C=1,N_D=2$ when $q < t$ or $N_C=2, N_D=1$ when $q < t$, and in both cases we can eliminate all but two: $B$ and $A \equiv \sum_i C_i + \sum_j D_j$. That $A$ is a valid path and has a well-defined phase is straightforward to show using 
\begin{eqnarray}
\bar{C}_i &=& -e^{ic_i} C_i = -e^{2\pi i \phi_{a, O6}} C_i^{\prime} \nonumber \\
\bar{D}_i &=& -e^{id_i} D_i = -e^{2\pi i \phi_{a, O6}} D_i^{\prime}
\end{eqnarray}
and \ref{CCPrimeSum}; we find
\beq
\bar{A} = -e^{2\pi i \phi_{a, O6}} A.
\eeq
The displacement associated with this is then
\beq
v_A = \frac{1}{\sqrt{2}} (v_a + v_{a'})= \sqrt{2} n_A L_a \cos \pi \phi_{a, O6}
\eeq

\subsubsection{$q<t$}

In this case, the matrix $M^{ab}$ is given by
\begin{eqnarray}
M^{AA} &=& i \frac{ \sin \pi (2\phi_{a, O6} + \theta) \tan \pi \phi_{a, O6}} {\sin \pi \theta} \nonumber \\
M^{AB} &=& e^{-i\pi \phi_{a, O6}}/2 \nonumber \\
M^{BA} &=& - e^{i\pi \phi_{a, O6}}/2 \nonumber \\
M^{BB} &=& 0 .
\end{eqnarray}
We also find 
\beq
\psi = -2\pi \phi_{a, O6}
\eeq
and thus $v_B$ is perpendicular to brane $a'$; we find
\beq
v_B = \sqrt{2} i \frac{e^{-\pi i \phi_{a, O6}}}{\cos \pi \phi_{a,O6}} \bigg( n_B  \frac{4\pi^2 T_2}{L_{a}} + y_B \bigg)
\eeq
where $y_B$ is the height of the smallest triangle $abO6$. 
The action is
\begin{multline}
S^{ab}_{cl} = \frac{1}{\Wh^{2}_A \Wh^{1}_B - \Wh^{1}_A \Wh^{2}_B }  \times\\
 \left(\begin{array}{cc} \bigg[\Wh^{1}_B \Wh^{2}_B  + iM^{AA} (\Wh^{1}_A \Wh^{2}_B + \Wh^{2}_A \Wh^{1}_B) \bigg] &  -e^{i\pi \phi_{a, O6}} M^{AA} \Wh^{1}_A \Wh^{2}_A \\ e^{-i\pi \phi_{a, O6}} M^{AA} \Wh^{1}_A \Wh^{2}_A & -\Wh^{1}_A \Wh^{2}_A  \end{array} \right) 
\end{multline}
which gives
\begin{eqnarray}\label{Actionqltt}
S_{cl} &=& \frac{1}{4\pi \ap} \frac{2}{\Wh^{2}_A \Wh^{1}_B - \Wh^{1}_A \Wh^{2}_B } \times   \\
&&\bigg[ \bigg( \Wh^{1}_B \Wh^{2}_B + iM^{AA} (\Wh^{1}_A \Wh^{2}_B + \Wh^{2}_A \Wh^{1}_B)\bigg)(n_A L_a \cos \pi \phi_{a,O6})^2 
 \nonumber\\
&& -\frac{\Wh^{1}_A \Wh^{2}_A}{\cos^2 \pi \phi} ( n_B  \frac{4\pi^2 T_2}{L_{a}} + y_B ) \bigg( n_B  \frac{4\pi^2 T_2}{L_{a}} + y_B + 2iM^{AA}  n_A L_a \cos^3 \pi \phi_{a,O6}\bigg) \bigg] \nonumber .
\end{eqnarray}
We also have the determinant
\beq
|W| = i e^{-3 \pi i \phi_{a, O6}} \bigg(\Wh^{1}_A \Wh^{2}_B - \Wh^{2}_A \Wh^{1}_B\bigg).
\eeq

For calculating the integrals $W^i_A$ when $\phi_{a, O6} \ne 0$ it is most expedient to use the identity
\beq
W^i_A = - W^i_{D_1} e^{-i\pi (\theta + \phi_{a, O6})} \frac{\sin \pi \theta}{\sin \pi (\phi_{a, O6})}
\eeq
from which one deduces that, in the limit $q\rightarrow 0$, that
\begin{eqnarray}
W^1_D &\rightarrow& \frac{\theta_1 (2\phi_{a, O6} i t)}{\theta_1^{\prime} (0)} e^{\pi i \theta} B(\theta, 1-\theta) \nonumber \\
W^1_A &\rightarrow& - \frac{\theta_1 (2\phi_{a, O6} i t)}{\theta_1^{\prime} (0)} e^{-\pi i \phi_{a, O6} } \frac{\pi}{\sin \pi ( \phi_{a, O6})} \nonumber \\
W^2_A &\rightarrow& - W^1_A \equiv -A
\end{eqnarray}
while it is also clear that $W^1_B \rightarrow W^2_B \equiv B \rightarrow \infty$. 

In this limit we find that the coefficient of the term quadratic in $n_A$ diverges, and thus the sum over $n_A$ is reduced to the zero mode: $\sum_{n_A} e^{-S(n_A,n_B)} \rightarrow e^{-S(0,n_B)}$. However, the coefficient of the quadratic term in $n_B$ reduces to zero, and we must Poisson resum, upon which the sum collapses to a single contribution:
\beq
\sum_{n_a, n_b} e^{-S(n_A,n_B)} \rightarrow \sqrt{\frac{|B|}{|A|}} \frac{\sqrt{\ap}}{2\pi} \frac{L_{a}\cos \pi \phi_{a,O6}}{ T_2} 
\label{ClassicalLimit}\eeq

\subsubsection{$q>t$}

In this case, the matrix $M^{ab}$ is given by
\begin{eqnarray}
\tilde{M}^{AA} &=& i \frac{ \sin \pi (2\phi_{a, O6} + \theta) \tan \pi (\theta + \phi_{a, O6}) } {\sin \pi \theta} \nonumber \\
\tilde{M}^{AB} &=& e^{-i\pi (\theta + \phi_{a, O6})}/2 \nonumber \\
\tilde{M}^{BA} &=& - e^{i\pi (\theta + \phi_{a, O6})}/2 \nonumber \\
\tilde{M}^{BB} &=& 0 .
\end{eqnarray}
We also find 
\beq
\psi = -2\pi \phi_{a, O6} - 2 \pi \theta,
\eeq
and thus $v_B$ is perpendicular to brane $b'$; we find
\beq
v_B = \sqrt{2} i \frac{e^{-\pi i (\theta + \phi_{a, O6})}}{\cos \pi (\phi_{a,O6} + \theta)} \bigg( n_B  \frac{4\pi^2 T_2}{L_{b}} + y_B \bigg).
\eeq
The action is
\begin{multline}
S^{ab}_{cl} = \frac{1}{\Wh^{2}_A \Wh^{1}_B - \Wh^{1}_A \Wh^{2}_B}  \times\\
 \left(\begin{array}{cc} \bigg[ \Wh^{1}_B \Wh^{2}_B  +i\tilde{M}^{AA} (\Wh^{1}_A \Wh^{2}_B + \Wh^{2}_A \Wh^{1}_B) \bigg] &  -e^{i\pi (\theta +\phi_{a, O6} )} \tilde{M}^{AA} \Wh^{1}_A \Wh^{2}_A \\ e^{-i\pi (\theta + \phi_{a, O6})} \tilde{M}^{AA} \Wh^{1}_A \Wh^{2}_A & -\Wh^{1}_A \Wh^{2}_A  \end{array} \right) 
\end{multline}
which gives the same action as equation (\ref{Actionqltt}) but with $M^{AA}$ replaced with $\tilde{M}^{AA}$ and $v_B$ modified. The determinant is
\beq
|W| = i e^{-3 \pi i \phi_{a, O6} - \pi i \theta} \bigg(\Wh^{1}_A \Wh^{2}_B - \Wh^{2}_A \Wh^{1}_B\bigg).
\eeq

For calculating the integrals $W^i_A$ when $\phi_{a, O6} \ne 0$ it is most expedient to use the identity
\beq
W^i_A = W^i_{C_2} e^{-i\phi_{a, O6}/2} \frac{\sin \pi \theta}{\sin \pi (\theta + \phi_{a, O6})}
\eeq
from which one deduces that, in the limit $q\rightarrow 0$, that
\begin{eqnarray}
W^1_{C_2} &\rightarrow& - \frac{\theta_1 (2 i (\theta + \phi_{a, O6}) t)}{\theta_1^{\prime} (0)} e^{-4\pi \theta t} B(\theta, 1-\theta) \nonumber \\
W^1_A &\rightarrow&  - e^{-i\pi\phi_{a, O6}} \frac{\theta_1 (2 i (\theta + \phi_{a, O6}) t)}{\theta_1^{\prime} (0)} e^{-4\pi \theta t} \frac{\pi}{\sin \pi (\theta + \phi_{a, O6})} \nonumber \\
W^2_A &\rightarrow& - W^1_A
\end{eqnarray}
while it is also clear that $W^1_B \rightarrow W^2_B \rightarrow \infty$.

\subsection{Quantum Part}

The quantum contribution for the M\"obius strip can be derived exactly as in \cite{Abel:2005qn} but with the new basis of cut differentials. The result is
\begin{multline}
\bra \prod_{i=1}^L {\sigma}_{\theta_i} (z_i) \ket_{\phi_{a, O6}} = |\tilde{W}|^{-1/2} \theta_{\phi_{a, O6}} (-Y)^{(L-M-1)/2} \ov{\theta_{-\phi_{a, O6}} (Y)}^{(M-1)/2} \\
\prod_{0< i< j}^{L-M} \theta_1 (z_i - z_j)^{1/2} \prod_{L-M< i< j}^{L} \theta_1 (z_i - z_j)^{1/2} \prod_{0< i< j}^{L} \theta_1 (z_i - z_j)^{-\frac{1}{2}[ 1 - \theta_i - \theta_j + 2\theta_i \theta_j]}
\end{multline}
written in the basis $\tilde{\omega}_i$. To transform to the basis $\omega_i$, we use \ref{convertbasesmob} and note that $|\tilde{W}| \rightarrow e^{4\pi i \phi_{a, O6} Y L} \prod_{i=1}^{L-M} e^{4\pi i \phi_{a, O6} z_i} \prod_{j=L-M+1}^{L} e^{-4\pi i \phi_{a, O6} z_j} |W|$ and $\theta_{\phi_{a, O6}} (-Y) \rightarrow e^{4\pi i \phi_{a, O6} Y} \theta_1 (Y + 2 \phi_{a, O6} i t)$ to give
\begin{multline}
\bra \prod_{i=1}^L {\sigma}_{\theta_i} (z_i) \ket_{\phi_{a, O6}} = |W|^{-1/2} \theta_1 (Y+2\phi_{a, O6} i t)^{(L-2)/2} e^{-4\pi i \phi_{a, O6} P} \\
\prod_{0< i< j}^{L-M} \theta_1 (z_i - z_j)^{1/2} \prod_{L-M< i< j}^{L} \theta_1 (z_i - z_j)^{1/2} \prod_{0< i< j}^{L} \theta_1 (z_i - z_j)^{-\frac{1}{2}[ 1 - \theta_i - \theta_j + 2\theta_i \theta_j]}
\end{multline}
where $P$ is as defined in \cite{Abel:2005qn}:
\beq
P \equiv \sum_{i=1}^L (1/2 - \theta_i) z_i .
\eeq

In the case of a two-point function with $z_1 = 0, z_2 = iq, \theta_1 = \theta, \theta_2 = 1-\theta, Y=\theta q i$ we have
\beq
\bra \sigma_\theta (0) \sigma_{1-\theta} (qi) \ket_{\phi_{a, O6}}^{qu} = |W|^{-1/2} e^{4\pi \phi_{a, O6} (\theta - 1/2) q} \left(\frac{\theta_1 (qi)}{\theta_1^{\prime} (0)}\right)^{-\theta (1-\theta)}
\label{Zqumob}\eeq

\subsection{Normalisation}

To normalise the two-point function we use the OPE
\beq
\bra \sigma_{\theta}^{ab} (z_1) \sigma_{1-\theta}^{ba}(z_2) \ket \sim (z_1 - z_2)^{-\theta (1-\theta)} C^{(aba)}_{\theta,1-\theta} M_{aa'}^X ,
\eeq
with the same OPE coefficients (\ref{ANNULUS:OPE}) as before, but now
 the partition function is
\beq
M_{aa'}^X = -i I_{aO6}\frac{\exp(4\pi (\phi_{a, O6})^2 t) \eta(it+1/2)}{\theta_1 ( 2 \phi_{a, O6} it, 1/2+it)} .
\eeq
Using equations (\ref{Zqumob}) and (\ref{ClassicalLimit}) we find for $q < t, q \rightarrow 0$
\beq
N (qi)^{-\theta (1-\theta)} e^{-\pi i /4} e^{3\pi i \phi_{a, O6}/2} \frac{1}{\sqrt{2|A||B|}} \sqrt{\frac{|B|}{|A|}} \frac{\sqrt{\ap}}{2\pi } \frac{L_{a}\cos \pi \phi_{a,O6}}{ T_2} = (qi)^{-\theta (1-\theta)} \frac{2\pi \sqrt{\ap}}{L_a^i} M_{aa'}^X
\eeq
and hence, using the identity (\ref{ANNULUS:LcLa}) but for the intersection between $a$ and $a'$ (with angle $2 \pi \phi_{a,O6}$) 
we find
\beq
N = \sqrt{2} \frac{e^{4\pi \phi_{a, O6}^2 t}}{ \eta^2 (it+1/2)}e^{\pi i /4 -3\pi i \phi_{a, O6}/2} 
\eeq

Following the same procedure for $q > t$ we factorise onto 
\beq
M_{bb'}^{X} = -i I_{bO6}\frac{\exp(4\pi (\phi_{a, O6}+\theta - 1)^2 t) \eta(it+1/2)}{\theta_1 ( 2 (\theta + \phi_{a, O6} - 1) it, 1/2+it)} 
\eeq
to obtain
\beq
\tilde{N} = N e^{-\pi i \theta/2}
\eeq

\subsubsection{$\phi_{a, O6}=0, -\theta$}

If either brane $a$ or $b$ is parallel to the orientifold plane ($\phi_{a, O6} = 0, -\theta$ respectively), then the partition function that we factorise onto contains worldsheet instantons. For $a$ parallel, for $q \rightarrow 0$ we factorise onto (\cite{Lust:2003ky}):
\beq
M_{aa'}^X = \frac{1}{\eta (it+1/2)^2} \sum_{r,s} e^{- t\frac{ 8\pi^3 \ap}{L_a^2} | r +   \frac{T}{\ap } s |^2} 
\eeq
 or the same for $M_{bb'}^X$ should that be the parallel brane, at the pole  $q \rightarrow 2t$. 
To see this, note that for $\phi_{a, O6}=0$ we have $M^{AA}=0$; but in addition, as in the case of an annulus diagram with a parallel brane, we find that $W^1_B = W^2_B \equiv B$ and $W^1_A = -W^2_A \equiv -i\hat{A}$ and thus the action (\ref{Actionqltt}) becomes
\beq
S_{cl} = \frac{1}{4\pi \ap} \bigg[ - \frac{B}{\hat{A}} (n_A L_a )^2 - \frac{\hat{A}}{B} (n_B  \frac{4\pi^2 T_2}{L_{a}} + y_B)^2 \bigg].
\eeq
and $|W| = 2i\hat{A}B$.
In the limit $q \rightarrow 0$ we find $B \rightarrow 1$, $\hat{A} \rightarrow 2t$, and so to show equivalence to the above we must perform a Poisson resummation on $n_A$ to obtain
\begin{eqnarray}
\bra \sigma_{\theta} (0) \sigma_{1-\theta} (iq) \ket &{}_{\quad\longrightarrow}^{\phi_{a, O6}\rightarrow  0}& \left(\frac{\theta_1 (iq)}{\theta_1^{\prime}(0)} \right)^{-\theta (1-\theta)} \frac{2\pi \sqrt{\ap}}{\eta(1/2+it)^2 L_{a}} \frac{1}{B} \nonumber \\
&&\times \sum_{\tilde{n}_A,n_B}e^{-\frac{4\pi^3 \ap}{L_{a}^2} \frac{\hat{A}}{B} [ (n_A )^2 + (n_B \frac{T_2}{\ap} + \frac{y_B L_{a}}{4\pi^2 \ap} )^2 ]   } 
\end{eqnarray}
which clearly reduces to the expected form in the limit. 

\subsection{Fermionic Correlators}

Calculation of fermionic correlators is straightforward using bosonised operators:
\beq
\bra \prod_{i=1}^L e^{i a_i H_i (z_i)} \ket_{\phi_{a, O6}, \nu} = e^{-4\pi i \phi_{a, O6} Q} \theta_{\nu} ( Q - 2  \phi_{a, O6} i t) \prod_{i < j} \theta_1 (z_i - z_j)^{a_i a_j}
\eeq
where $\nu$ indicates the spin structure and 
\beq
Q \equiv \sum_i a_i z_i .
\eeq
To normalise, we require the fermionic partition function
\beq
M_{aa'}^{\psi} = -i \frac{\theta_{\nu} ( 2 \phi_{a, O6} it, 1/2+it)}{\exp(4\pi (\phi_{a, O6})^2 t)\eta(it+1/2) } .
\eeq
and thus we must multiply by $i \eta^{-1} (it+1/2) e^{-4\pi \phi_{a, O6}^2 t}$.

For the two-point function with $a_1 = \theta - 1, a_2 = 1 - \theta$ we have
\begin{eqnarray}
\bra e^{i(\theta-1)H (0)} e^{-i(\theta-1)H(qi)} \ket = e^{4\pi  \phi_{a, O6} (1-\theta)q}i \theta_1 (qi)^{-(1-\theta)^2} \frac{\theta_{\nu} ( 2 \phi_{a, O6} i t - (1-\theta) q i ) }{e^{4\pi \phi_{a, O6}^2 t}\eta (it+1/2)} 
\end{eqnarray}

\subsection{Full Two-Point M\"obius Amplitude}

We now assemble the above machinery to compute the two-point function for the M\"obius strip in $N=1$ supersymmetric sectors. The contribution to sectors with more supersymmetry can be obtained from the below by setting some angles to zero. We use the same vertex operators (\ref{ANNULUS:Vertices}) as the previous section, for states at an intersection with angles $\theta^{\kappa}$ where $\sum \theta^{\kappa} = 2$, but we define
\beq
z_1 = 0, \qquad z_2 = iq
\eeq
in accordance with the method outlined in this section (but in contrast to that used in the previous one). 
Here we also have the angle between brane $a$ and the orientifold plane $\phi_{a, O6}^{\kappa}$. Here we take $\sum_{\kappa} \phi_{a, O6}^{\kappa} = 2$, although summing to zero is entirely equivalent for these.

We thus write
\begin{eqnarray}
\mathcal{M} &\equiv& \bra C_{ab} (k) \ov{C}_{ab} (-k)\ket \nonumber \\ &=& 4(\ap)^2 k^2G_{C_{ab},\ov{C}_{ab}}\int_0^{\infty} \frac{dt}{(8\pi^2 \ap t)^2} \frac{1}{\eta (it+1/2)^3} \int_0^{2t} dq (2\sqrt{2}) \nonumber \\&& \times
\chi(qi) \left(\frac{\theta_1(iq)}{\theta_1^{\prime}(0)}\right)^{-2} 
e^{4\pi q} \nonumber \\ && \times \sum_{\nu} \delta_{\nu} \theta_\nu (qi)  
\prod_\ka \frac{\theta_\nu (2\phi_{a, O6}^{\ka} it + (\theta^{\ka} -1)qi )}{\eta(it+1/2)^3} 
|W^{\ka}|^{-1/2} \sum_{n^{\ka}_A,n_B^{\ka}} e^{-S^{\ka}} 
\end{eqnarray}
which after summation over spin structures becomes
\begin{multline}
\mathcal{M} = \frac{k^2}{16\pi^2} G_{C_{ab},\ov{C}_{ab}}\int_0^{\infty} \frac{dt}{ t^2} \frac{1}{\eta(it+1/2)^6} \int_0^{2t} dq  \theta_1(iq)^{-1} e^{4\pi q}\chi(qi)  \\
\times \prod_\ka \theta_1(2\phi_{a, O6}^{\ka} it + (\theta^{\ka} -1)qi ) 
|W^{\ka}/2|^{-1/2} \sum_{n^{\ka}_A,n_B^{\ka}} e^{-S^{\ka}}  .
\end{multline}
In the same way as the previous section, but after rescaling $t \rightarrow t/4$ to match the modular parameter of the M\"obius strip to the annulus we can find the pole behaviour 
\beq
\mathcal{M} = -4 \rho_{\Omega R} \frac{G_{C_{ab},\ov{C}_{ab}}}{32\pi^3} tr(\lambda_{ab} \lambda_{ab}^{\dagger})\bigg[ \frac{(2\pi \sqrt{\ap})^3}{2\ap} \bigg( \frac{ I_{ac}}{L_a} + \frac{ I_{bc}}{L_b}\bigg) \int_0^{\infty} \frac{dt}{t^2} + \mathrm{finite}\bigg]
\eeq
where, in the same way as in section 2, we have $\gamma_{\Omega R}^{-1} \gamma_{\Omega R}^{T}= \rho_{\Omega R} {\bf 1} = \pm {\bf 1}$. This is then exactly the correct contribution to cancel the poles in the annulus diagrams.

\section{Conclusions}

We have calculated the one loop K\"ahler metric for chiral fields on branes in both branes at orbifold fixed points and intersecting brane models, and in so doing completed the set of techniques for calculating $D6$-brane boundary-changing operator amplitudes in toroidal orientifold models. The two types of calculations are in stark contrast, due to the presence of the boundary changing operators in the second case, although they both reproduce the field theory expectations and both contain closed string tadpoles that must be subtracted. In addition, the first computation involves no moduli dependence, and so we expect that the corrections given have a universal quality independent of the geometry. The intersecting branes, on the other hand, have K\"ahler and brane modulus dependence through the worldsheet instantons, and so are sensitive to the whole compact space. It is nevertheless possible to use these latter computations in various limiting cases; for example intersecting branes can be used as a toy model for D-term supersymmetry breaking, but we postpone such calculations for further work.

\section{Acknowledgments}

Work supported in part by the French ANR contracts
BLAN05-0079-01 and \\
PHYS@COL\&COS. M.~D.~G. is supported by a CNRS contract.

\appendix

\section{Theta Functions}
\label{App:A}

The theta functions are defined as
\beq
\tab{a}{b} (z,\tau) = \sum_{n = -\infty}^{\infty} \mathrm{exp} \bigg[ \pi i (n+a)^2 \tau + 2 \pi i (n+a)(z+b) \bigg]
\eeq
and have periodicies
\begin{eqnarray}
\theta \left[ \begin{array}{c} a \\ b \end{array} \right] (z+m;\tau) &=& \exp (2\pi i am) \theta \left[ \begin{array}{c} a \\ b \end{array} \right] (z;\tau) \\
\theta \left[ \begin{array}{c} a \\ b \end{array} \right] (z+m\tau;\tau) &=& \exp (-2\pi i bm) \exp (-\pi i m^2 \tau - 2 \pi i mz)\theta \left[ \begin{array}{c} a \\ b \end{array} \right] (z;\tau) \nonumber 
\end{eqnarray}

The modular transformation of $\theta_1(z,\tau) \equiv \tab{\frac{1}{2}}{\frac{1}{2}} (z,\tau)$ is
\begin{eqnarray}
\theta_1(z, \tau) &=& i (-i\tau)^{-1/2} \exp(-\pi i z^2/\tau) \theta_1 (z/\tau,-1/\tau) \nonumber \\
&=& \exp(-\pi i /4) \theta_1 (z,\tau + 1) \nonumber \\
\theta_1 (z,1/2 + it) &=& \sqrt{\frac{i}{2t}} \exp ( -\pi z^2 / t )\theta_1 ( \frac{i z}{2t},\frac{i}{4t} -1/2)
\end{eqnarray}

Some additional identities used in the text are presented below.
\beq
\tab{\alpha + c}{\beta + d} (z,\tau) = \exp [ 2\pi i c (z + d + \alpha) + c^2 \pi i \tau] \tab{\alpha}{\beta} (z+ c \tau + d,\tau)
\label{A:thetapluscd}\eeq

\beq
\frac{\theta_1^{\prime} (0) \theta_1 (a+b)}{\theta_1 (a) \theta_1 (b)} = \pi \cot (\pi a) + \pi \cot (\pi b) + 4\pi \sum_{m, n = 1}^{\infty} e^{2\pi m n i \tau} \sin (2\pi m a + 2 \pi n b )
\label{A:thetaaplusb}\eeq

\beq
\frac{\theta_1^{\prime} (z)}{\theta_1(z)} = \pi \cot \pi z + 4 \pi \sum_{m,n=1}^{\infty} e^{2\pi m n i \tau} \sin (2\pi m z)
\label{A:thetaprimeovertheta}\eeq

\end{document}